\documentclass[prapplied, reprint, superscriptaddress, amsmath, amssymb, aps]{revtex4-1}

\pdfoutput=1

\usepackage{graphicx}% Include figure files
\usepackage{dcolumn}% Align table columns on decimal point
\usepackage{bm}
\usepackage{placeins}
\usepackage{xcolor}
\usepackage{soul}
\usepackage[resetlabels]{multibib}
\usepackage{lineno}
\usepackage{natbib}
\usepackage{siunitx}
\usepackage{multirow}

\newcommand{\eqlab}[1]{\label{eq:#1}}
\renewcommand{\eqref}[1]{Eq.~(\ref{eq:#1})}

\newcommand{\figref}[1]{Fig.~\ref{fig:#1}}

\begin{document}

\preprint{APS/123-QED}

%%%%%%%%%%%%%%%%%%%%%%%%%%%%%%%%%%%%%%%%%%%%%%%%%%%%%%%%%%%%%%%%%%%%%%%%%%%%%%%%%%%%%%%%%%%%%%%%%
% TITLE AND AUTHOR INFO
%%%%%%%%%%%%%%%%%%%%%%%%%%%%%%%%%%%%%%%%%%%%%%%%%%%%%%%%%%%%%%%%%%%%%%%%%%%%%%%%%%%%%%%%%%%%%%%%%

\title{Terahertz light sources by electronic-oscillator-driven \\ second harmonic generation in extreme-confinement cavities}

\author{Lamia Ateshian}
\altaffiliation{These authors contributed equally to this work.}
\author{Hyeongrak Choi}
\altaffiliation{These authors contributed equally to this work.}
\author{Mikkel Heuck}
\author{Dirk Englund}
\email{englund@mit.edu}
\affiliation{%
 Research Laboratory of Electronics, Massachusetts Institute of Technology, Cambridge, Massachusetts 02139, USA
}%
\affiliation{%
 Department of Electrical Engineering and Computer Science, Massachusetts Institute of Technology, Cambridge, Massachusetts 02139, USA
}%

\date{\today}

\begin{abstract}
The majority of sources of coherent optical radiation rely on laser oscillators driven by population inversion. Despite their technological importance in communications, medicine, industry, and other fields, it remains a challenge to access the spectral range of 0.1-10 THz (the ``terahertz gap''), a frequency band for applications ranging from spectroscopy to security and high-speed wireless communications. Here, we propose a way to produce coherent radiation spanning the THz gap by efficient second-harmonic generation (SHG) in low-loss dielectric structures, starting from technologically mature electronic oscillators (EOs) in the $\sim$100 GHz range. To achieve this goal, we introduce hybrid THz-band dielectric cavity designs that combine (1) extreme field concentration in high-quality-factor resonators with (2) nonlinear materials enhanced by phonon resonances. We theoretically predict conversion efficiencies of $>10^3\%$/W and the potential to bridge the THz gap with 1 W of input power. This approach enables efficient, cascaded parametric frequency converters, representing a new generation of light sources extensible into the mid-IR spectrum and beyond.
\end{abstract}

\maketitle

%%%%%%%%%%%%%%%%%%%%%%%%%%%%%%%%%%%%%%%%%%%%%%%%%%%%%%%%%%%%%%%%%%%%%%%%%%%%%%%%%%%%%%%%%%%%%%%%%
% INTRODUCTION
%%%%%%%%%%%%%%%%%%%%%%%%%%%%%%%%%%%%%%%%%%%%%%%%%%%%%%%%%%%%%%%%%%%%%%%%%%%%%%%%%%%%%%%%%%%%%%%%%

\section{Introduction}
\begin{figure*}[hbt!]
    \centering
    \includegraphics[width=\textwidth]{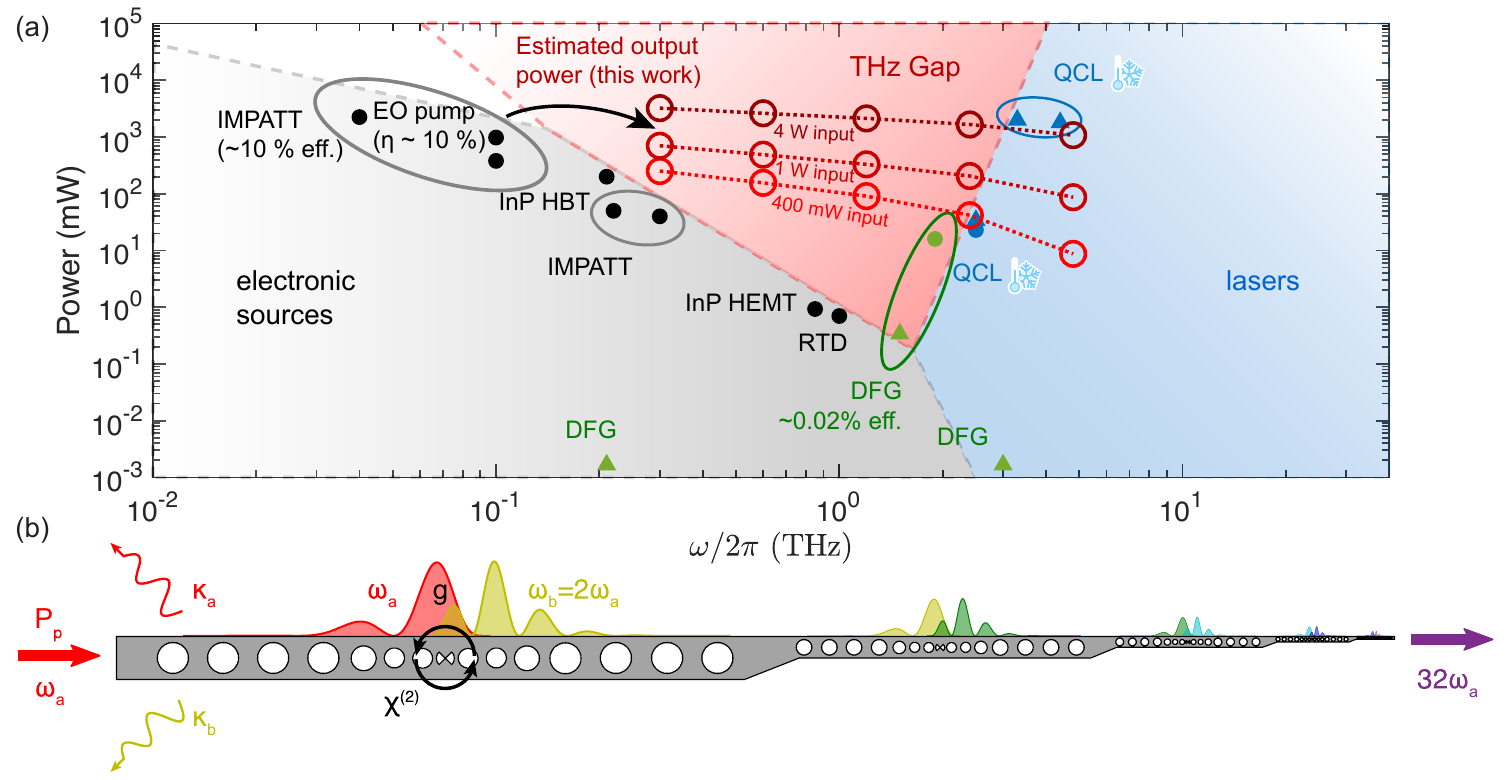} % updated 8.14.21 with ring data
    \caption{Terahertz sources and proposed work. (a) Selected terahertz source technologies~\cite{tonouchi2007cutting} based on electronics (black), lasers (blue), and difference frequency generation (DFG) in nonlinear crystals (green)
    ~\cite{midford1979millimeter,niehaus1973double,urteaga2017inp,leong20150,izumi2019resonant,jin2020phase,luo2020distributed,li2017multi,bondaz2019generation,yan2017high,petersen2011efficient}). 
    % ~\cite{jin2020phase,luo2020distributed,li2017multi,leong20150,deal2016inp,izumi2019resonant,urteaga2017inp,cao2019w,bondaz2019generation,wu2019generation,kasagi2019large}). 
    ``Terahertz gap'' is shaded in red; available output power in 100 GHz - 10 THz range is limited.  Approximate wall-plug efficiency is indicated where available. Circles denote CW power and triangles denote average pulse power. 
    %HBT: heterojunction bipolar transistor. HEMT: high-electron-mobility transistor. RTD: resonant tunneling diode. QCL: quantum-cascade laser. 
    Our proposed device     (results for ring cavity shown here) starts with an EO input at $\sim$1 W, $\sim$100 GHz (circled) to generate output power crossing the THz gap by cascaded frequency doubling. 
    For comparison, output power for 400 mW and 4 W input power are also shown. In this calculation, we included the radiation loss of the SH cavity mode and neglected direct radiation loss from the nonlinear polarization field into other modes at the SH frequency~\cite{suppConversionEff}. %(see Supplementary Section 1A).
    %Open circles demonstrate the device power with higher-power input sources. 
    (b) Schematic of spectrum-spanning nonlinear frequency synthesis approach in PhC cavities. The fundamental mode at $\omega_a$ is coupled to the second-harmonic mode at $\omega_b = 2\omega_a$ in the first cavity with coupling rate $g$. The output at $\omega_b$ couples to the next cavity, cascading in a sequence of frequency doubling steps. After $N$ steps the final output is at $2^N\omega_a$ (here $N=5$).} 
    \label{fig:cavity}
\end{figure*}

There are fundamental differences in how frequency-stable electromagnetic (EM) radiation is generated at frequencies $\omega/2\pi \ll 1$ THz and $\omega/2\pi \gg 1$ THz. The majority of sub-THz sources rely on electronic oscillators (EOs) or frequency multipliers. Far above $\sim$10 THz, sources use gain media based on population inversion. Separating these frequency regimes is the ``THz gap,'' commonly defined as 0.1 - 10 THz, in which efficient, compact, and room-temperature EM sources have been notoriously challenging to build. However, the abundance of opportunities in the THz spectrum for a range of applications -- from molecular spectroscopy to remote sensing, navigation, and wireless communication~\cite{Lewis2014,Han2019,Sengupta2018} -- motivates the development of more efficient sources in this band. Within the THz gap (\figref{cavity}(a)) sources based on electronic methods are possible through nonlinear electrical frequency-multipliers and high-frequency oscillators, but their operation becomes inefficient above the maximum operational frequency of transistors $f_{max}\sim 100 - 300$ GHz~\cite{Han2019, aghasi2020terahertz}. On the other hand, THz sources derived from population inversion have low efficiencies (e.g. $0.02\%$ for optical DFG~\cite{petersen2011efficient}), require cryogenic cooling (e.g. quantum-cascade lasers~\cite{Fujita2018}), or rely on expensive and bulk ultra-fast lasers~\cite{perkowitz2020navigating}. 

Here, we introduce an approach based on extreme field concentration with high quality factor cavities~\cite{choi2017self,Hu2016} that achieves frequency conversion from $\sim 100$ GHz (microwave) into the THz domain with efficiencies exceeding $10^3\,\%/$W. Our approach opens the door to efficient, phase-stable synthesis of electromagnetic radiation bridging the terahertz gap.

As indicated in ~\figref{cavity}(a), we consider an electronic source at the low end of the THz gap that drives the first stage of second-harmonic generation (SHG) in a dielectric cavity. This cavity is made of a low-loss dielectric material that creates a deep sub-wavelength region of high electric energy density inside a $\chi^{(2)}$ nonlinear material. After the first conversion step ($\omega_a$ to $\omega_b = 2\omega_a$), the output is injected into the next cavity. After cascading $N$ cavities, the output frequency is $2^N \omega_a$. The absence of a gain medium eliminates numerous challenges such as quantum noises (spontaneous emission) and technical noises (drive current noise, relaxation oscillations, vibrations, etc), promising phase-stability inherited from the electronic seed oscillator. 

The efficiency of the SHG approach increases with the pump power and is only limited by the dielectric breakdown of materials. Because the SHG is ``parametric'' (which does not dissipate power), our SHG approach mitigates the sharp efficiency drop in electronic sources above $f_{max}$ due to ohmic losses and other parasitic losses of nonlinear reactances~\cite{aghasi2020terahertz}. Figure \ref{fig:cavity}(a) plots the estimates of the resulting output power at each stage of the SHG cascade with a 400 mW, 1 W, and 4 W input. Further advances in EO output power, cavity designs, and nonlinear materials should further increase the efficiency spanning the entire THz gap.

Figure \ref{fig:cavity}(b) illustrates the SHG cascade using a doubly-resonant photonic crystal (PhC) cavity. Pumping mode $a$ with a power $P_p$ generates a field in mode $b$ with efficiency $\eta_\text{SHG}$~\cite{suppConversionEff}. %(see Supplementary Section 1)
\begin{align}\label{eq:eta_shg}
     \eta_\text{SHG} \equiv \frac{P_\text{SHG}}{P_p^2} = \frac{64}{\hbar\omega_a^4}g^2Q_a^2Q_b\eta_c,
\end{align}
where $\eta_c$ is the input-output coupling efficiency, $Q_{a}$ and $Q_{b}$ are the quality factors of the fundamental (FD) and second-harmonic (SH) modes, respectively, and $g$ is the nonlinear coupling rate, given by
\begin{align}\eqlab{g def}
    g &= \chi^{(2)}_\text{eff}\sqrt{\frac{\hbar\omega_a^2\omega_b}{\epsilon_0}}\frac{\tilde{\beta}}{\sqrt{(\lambda_a/n_a)^3}},
\end{align}
where $\chi_\text{eff}^{(2)}$ is the effective second-order nonlinear coefficient, $n_a$ is the refractive index of the nonlinear material at $\omega_a$, and $\tilde{\beta}$ is the SHG mode overlap between the FD and SH modes, normalized to the wavelength in the nonlinear medium. 

Considering $\omega_b, 1/\lambda_a \propto \omega_a$, the coupling rate $g$ is proportional to $\omega_a^3$, and thus,  $\eta_\text{SHG}\propto \omega_a^2$. The dependence of the conversion efficiency on the frequency squared poses a technical challenge in THz SHG compared to its optical counterpart. We will show that this scaling can be overcome with a combination of (1) materials with large nonlinear coefficient derived from phonon resonances and (2) cavity designs with strong field confinement.

%%%%%%%%%%%%%%%%%%%%%%%%%%%%%%%%%%%%%%%%%%%%%%%%%%%%%%%%%%%%%%%%%%%%%%%%%%%%%%%%%%%%%%%%%%%%%%%%%
% NONLINEAR MATERIAL ANALYSIS
%%%%%%%%%%%%%%%%%%%%%%%%%%%%%%%%%%%%%%%%%%%%%%%%%%%%%%%%%%%%%%%%%%%%%%%%%%%%%%%%%%%%%%%%%%%%%%%%%

\section{Nonlinear material analysis}
We make use of large second-order nonlinear susceptibilities derived from transverse optical (TO) phonon resonances. Transverse optical phonons can be driven by EM waves, resulting in large linear susceptibilities. The nonlinear susceptibilities are higher on resonance because they are proportional to the linear susceptibilities at the frequency components of interest~\cite{boyd2003nonlinear}. Phonon resonance frequencies lie at several THz for the crystals we consider here: GaAs, GaP, and ZnTe (zinc-blendes, class $\bar{4}3m$); and LiTaO$_3$ and LiNbO$_3$ (ferroelectrics, class $3m$).

Since experimental data on THz nonlinear optical susceptibilities in materials are limited, we rely on theoretical models supported by the existing data. The Faust-Henry model \cite{faust1968dispersion,flytzanis1969second} is used for zinc-blende crystals and an extension of Miller’s rule \cite{garrett1968nonlinear} for ferroelectric crystals. In both models, nonlinear coefficients are expressed by products of Lorentzian oscillators~\cite{suppNonlinSuscep}. %(Details of the calculations of nonlinear susceptibilities can be found in Supplementary Section 2.)

\begin{figure}[ht!]
    \centering
    \includegraphics[width=\columnwidth]{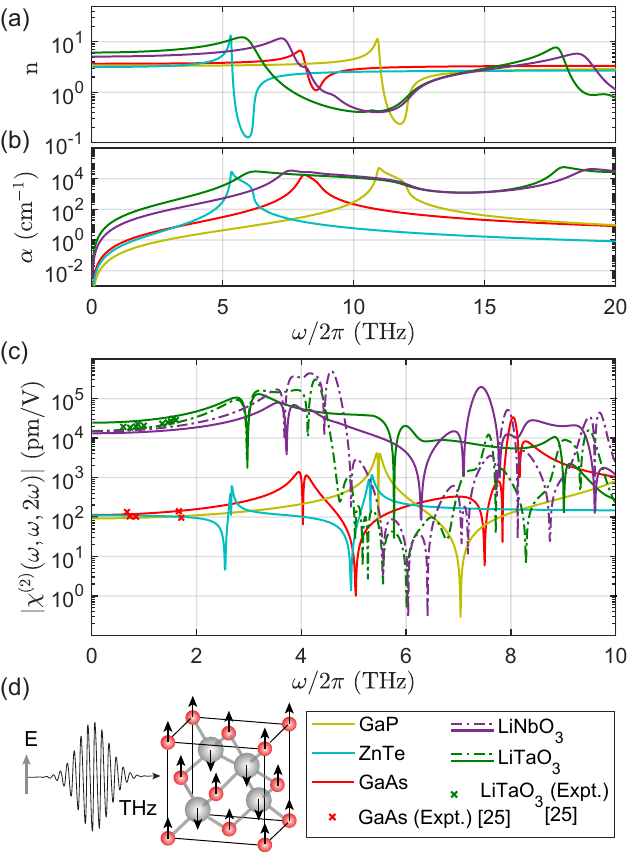} % updated fonts 6.14.21
    \caption{Susceptibilities predicted from the theoretical models along with experimental data. (a) Refractive index ($n_e$ for LiTaO$_3$ and LiNbO$_3$). (b) Absorption coefficient. (c) Second-order susceptibility, $|\chi^{\scriptscriptstyle (2)}_{jkl}(\omega, \omega, 2\omega)|$. Solid (using contracted notation): $\chi^{(2)}_{33}$ (LiTaO$_3$, LiNbO$_3$) or $\chi^{(2)}_{14}$ (GaAs, GaP, ZnTe). Dashed: $\chi^{(2)}_{31}$ (LiTaO$_3$, LiNbO$_3$).
    % Tensor component $\chi^{(2)}_{14}$ is shown for GaAs, GaP, ZnTe; $\chi^{(2)}_{33}$ (solid) and $\chi^{(2)}_{31}$ (dashed) for LiNbO$_3$ and LiTaO$_3$. Points indicate available experimental data for GaAs and LiTaO$_3$ ($\chi^{(2)}_{31}$) from~\cite{mayer1986far}. 
    $\chi^{(2)}$ is plotted over half the frequency range of $n$ and $\alpha$ since SHG from $\omega_a$ to $2\omega_a$ depends on $n$ and $\alpha$ at both frequencies. Crosses indicate available experimental data~\cite{mayer1986far}.
    (d) Illustration of transverse optical phonon mode excited by an incident THz field. When either the fundamental or second harmonic mode lies near the TO phonon frequency, the nonlinear susceptibility rapidly rises due to the resonance. The first maximum appears at half the lowest TO phonon frequency, for example, at $\sim 4.0$ THz in GaAs. A second peak occurs at $\omega_{TO} \sim 8.0$ THz, coinciding with the resonant features in the index and absorption coefficient in (a) and (b).
    % Optical phonon excites out-of-phase oscillation between atoms (a). Comparison of THz refractive indices (b) absorption coefficients (c), and second order nonlinear susceptibilities (d) of five materials. Tensor component of $\chi^{(2)}$: 14 shown for GaAs, GaP, ZnTe and 33 shown for LiNbO$_3$, LiTaO$_3$. 
     }
    \label{fig:thz-params}
\end{figure}

The dispersion of the linear susceptibility $\chi^{(1)} = \epsilon - 1 = (n-\frac{ic\alpha}{2\omega})^2 - 1$ (where $\epsilon$ is the relative permittivity) is calculated by the damped oscillator model. The refractive index, $n$, and absorption coefficient, $\alpha$, are plotted in Figs.~\ref{fig:thz-params}(a) and (b). Parameters used in the calculations (e.g., TO phonon frequencies, damping constants, and oscillator strengths) can be found in Table S3 in the Supplemental~\cite{suppNonlinSuscep} and Refs.~\cite{dekorsy2003infrared, mayer1986far,faust1966mixing, barker1968dielectric,carnio2020extensive, hattori1973indices,barker1970infrared,barker1967dielectric}. We note that although the theoretical prediction of the absorption coefficient goes to 0 as $\omega \rightarrow 0$, experiments using THz time-domain spectroscopy show appreciable absorption at low frequencies due to other dissipation processes not included in this model ~\cite{kojima2003dielectric,kojima2014broadband,grischkowsky1990far,mayer1986far}. 

\figref{thz-params}(c) plots $|\chi^{(2)}(\omega,\omega,2\omega)|$ as a function of the fundamental frequency from 0 to 10 THz. In the zinc-blende materials, the nonzero components of the nonlinear tensor $\chi^{(2)}_{14}$, $\chi^{(2)}_{25}$, and $\chi^{(2)}_{36}$ are all equal. For the ferroelectric crystals, $\chi^{(2)}_{33}$ and $\chi^{(2)}_{31}$ are shown. 

Our calculations indicate that these phonon resonances result in remarkably high THz nonlinear susceptibilities for LiTaO$_3$ and LiNbO$_3$ of over $10^4$ pm/V, exceeding their values in the optical range by around three orders of magnitude ($\chi^{(2)}_{33}=-40$ and $-60$ pm/V, respectively~\cite{boyd1971microwave,boyd2003nonlinear}). GaAs, GaP, and ZnTe also show an order of magnitude increase relative to their optical counterparts ($\chi^{(2)}_{14} = 268, 156$, and $139$ pm/V, respectively~\cite{boyd1971microwave,carnio2020extensive}).

Early measurements of $\chi^{(2)}(\omega,\omega,2\omega)$ at THz frequencies in GaAs and LiTaO$_3$~\cite{mayer1986far} provide a few experimental data points that agree with our calculated predictions. Follow-up work in GaAs additionally observed the resonant enhancement of SHG at half the phonon energy~\cite{dekorsy2003infrared}. 

The large $\chi^{(2)}$ coefficients are accompanied by high absorption losses, as shown in ~\figref{thz-params}(b). As such, standard cavity designs using a single material do not benefit from the $\chi^{(2)}$ due to the strong linear absorption. To overcome the loss, we introduce hybrid cavity designs in which the nonlinear materials are embedded in a low-loss dielectric, e.g., high-resistivity Si~\cite{suppSiLoss}. %(see Supplementary Section 3F.) 
This allows us to take advantage of the $\chi^{(2)}$ while minimizing material losses.

%%%%%%%%%%%%%%%%%%%%%%%%%%%%%%%%%%%%%%%%%%%%%%%%%%%%%%%%%%%%%%%%%%%%%%%%%%%%%%%%%%%%%%%%%%%%%%%%%
% CAVITY DESIGNS
%%%%%%%%%%%%%%%%%%%%%%%%%%%%%%%%%%%%%%%%%%%%%%%%%%%%%%%%%%%%%%%%%%%%%%%%%%%%%%%%%%%%%%%%%%%%%%%%%

\begin{figure*}[hbt!]
    \centering
    \includegraphics[width=\textwidth]{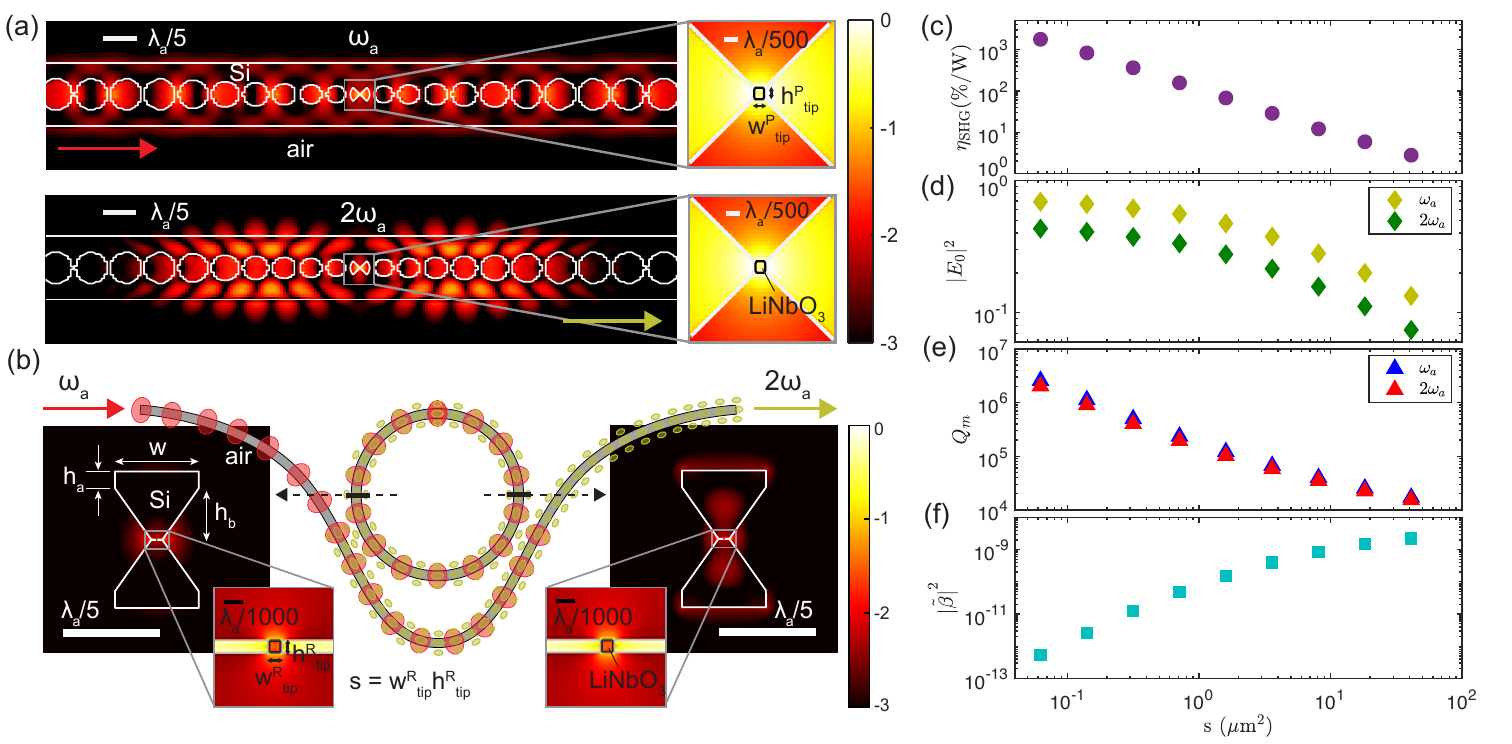} % updated 6/29/21
    \caption{Cavity modes and scaling behavior. (a) Top-view of the field profiles ($\log_{10} \big| \frac{\mathbf{E}}{\mathbf{E}_{max}} \big|^2$) of PhC cavity modes (TE) at $\omega_a = 331$ GHz (top) and $\omega_b = 662$ GHz (bottom). The thickness of the cavity (out of plane) is $\sim 0.153 \lambda_a$. Scale-bars are $\lambda_a/5$ ($\lambda_a/500$ for close-ups). Note that we only show 10 holes on each side for the visibility while the whole structure has 32 holes on each side. (b) Side-view of the field profiles in the ring for TM (vertically polarized) modes at $\omega_a = 350$ GHz (left), $\omega_b = 700$ GHz (right). Here, LiNbO$_3$ is used as the nonlinear material and Si as the host material for both the PhC and ring cavities. (c) The SHG efficiencies with cross-sectional area $s$ of nonlinear material in the ring resonator, for conversion from 350 GHz to 700 GHz. (d) Normalized electric field square in the tip $|E_0|^2 = \left(\frac{\lambda_a}{n_a}\right)^3 \frac{\epsilon_\text{tip}|\mathbf{E}_\text{tip}|^2}{\int \epsilon(\mathbf{r}) |\mathbf{E}(\mathbf{r})|^2 d^3\mathbf{r} }$ where $|\mathbf{E}_\text{tip}|$ is the average field amplitude in the nonlinear material. (e) Material quality factor. (f) Magnitude squared of normalized nonlinear overlap $|\tilde{\beta}|^2$, calculated for a ring radius of 3.6$\lambda_a$.}
    \label{fig:modes}
\end{figure*}

\section{Cavity Designs}
Dielectric cavities with strong field confinement and high quality factors were recently studied~\cite{choi2017self,Hu2016,choi2019cascaded,lu2017ultrahigh} and experimentally demonstrated~\cite{Hu2018}. These designs introduce tip structures at the anti-node of a mode resulting in extreme field enhancement at the dielectric tip. 

Here, we apply this field concentration principle to the THz regime to design hybrid-material PhC and ring cavities with large SHG conversion efficiency. Both cavity types have unique benefits and challenges; PhC cavities have smaller mode volume and do not need phase matching; ring cavities require phase matching but the radiation $Q$ is less sensitive to imperfections and easier to couple to a waveguide without modification.
%PhC cavities have higher efficiency from strong confinement and insensitivity to phase-matching due to the small size of the nonlinear material in the longitudinal direction. Ring cavities require phase-matching, but generally have higher quality factor and waveguide-coupling is easily engineered. 
% in- and out-coupling to waveguides is more accessible without cavity modification. 

\figref{modes}(a) shows the FD (top) and SH mode (bottom) of our nanobeam PhC cavity design (at the middle cross-section). The holes in the rectangular waveguide form a Bragg reflector. The tapering of the hole radii creates defect modes in the bandgap. The dielectric field concentration tips are introduced in the center hole~\cite{choi2019cascaded}. A rectangular nanowire made of nonlinear medium sits between the two concentric tips where the electric field is strongest (inset). One can position nanowires using a transfer setup \cite{notomi2020nanowire}. 

\figref{modes}(b) shows a top view of our ring cavity (middle) and cross sectional profiles of the FD (left) and SH mode (right). Interferometric coupling~\cite{Madsen1999} can be used for tuning of the coupling Q factor, $Q_{a(b)}^c$, that optimizes the conversion efficiency~\cite{suppConversionEff}. %(see Supplementary Section 1). 
This type of cavity can be fabricated by combining two separately fabricated ring resonators with angled cross sections, one with a thin nonlinear medium on top of the silicon.

We use the $\chi^{(2)}_{33}$ component to couple the fundamental transverse magnetic (TM) mode at $\omega_a$ and a higher order TM mode at $2\omega_a$. To couple the two TM modes by the $\chi^{(2)}_{14}$ component in the zinc-blende materials, the dominant electric field component should be aligned to the [111] crystal axis~\cite{buckley2013second}, in which case $\chi^{(2)}_\text{eff} = \frac{2}{\sqrt{3}}\chi^{(2)}_{14}$. 
The ring design achieves phase matching by the condition on the angular wavenumber, $m_b-2m_a = 0$.
Selecting the radius such that the ring is resonant at the frequencies $\omega_a$ and $2\omega_a$ for which the effective indices of the waveguide are equal fulfills the energy matching condition~\cite{suppCavityDesigns}. %Supplementary Sections 3A-D provide more details on the cavity designs.

Figs.~\ref{fig:modes}(c-f) illustrate how our hybrid cavity designs enable large SHG conversion efficiencies using the ring cavity as an example. Here we assume the cavity host material (Si) to be lossless~\cite{suppSiLoss}. %(see Supplementary Section 3F for a discussion of losses in Si). %and a radiation-limited quality factor of the ring of  $Q_{a(b)}^{r} = 10^6$. 
The insets in~\figref{modes}(a) and (b) show that the nonlinear material is embedded only near the tip where both cavity modes have a high electric energy density. The material quality factor $Q_{a(b)}^m$ is proportional to the inverse of the
%The loss rate corresponding to material loss, $\kappa_{a(b)}^m = \omega_{a(b)}/2Q_{a(b)}^m$, is reduced by the 
fraction of energy in the $\chi^{(2)}$ material~\cite{suppMatAbsorption}. %(see Supplementary Section 3E).
% , which depends on the square of the electric field in the tip gap. 

As shown in \figref{modes}(c), which plots the conversion efficiency $\eta_{\rm{SHG}}$ as a function of the cross-sectional area $s$ of the nonlinear medium, the efficiency increases as the size of the nonlinear medium is reduced. \figref{modes}(d) plots the normalized electric field square $|E_0|^2$ in the nonlinear material. As $|E_0|^2$ increases with decreasing nonlinear material volume, $Q_{a(b)}^m$ also increases, as plotted in~\figref{modes}(e). The nonlinear mode overlap $|\tilde{\beta}|$ decreases with nonlinear material volume (see~\figref{modes}(f)), but $\eta_{\rm{SHG}}$ increases since it is proportional to the quality factor cubed. This increase saturates when the total quality factor is dominated by radiation loss. 

%%%%%%%%%%%%%%%%%%%%%%%%%%%%%%%%%%%%%%%%%%%%%%%%%%%%%%%%%%%%%%%%%%%%%%%%%%%%%%%%%%%%%%%%%%%%%%%%%
% CONVERSION EFFICIENCIES
%%%%%%%%%%%%%%%%%%%%%%%%%%%%%%%%%%%%%%%%%%%%%%%%%%%%%%%%%%%%%%%%%%%%%%%%%%%%%%%%%%%%%%%%%%%%%%%%%

\section{Conversion Efficiencies}
\begin{figure} [!hbtp]
    \centering
    \includegraphics[width=\columnwidth]{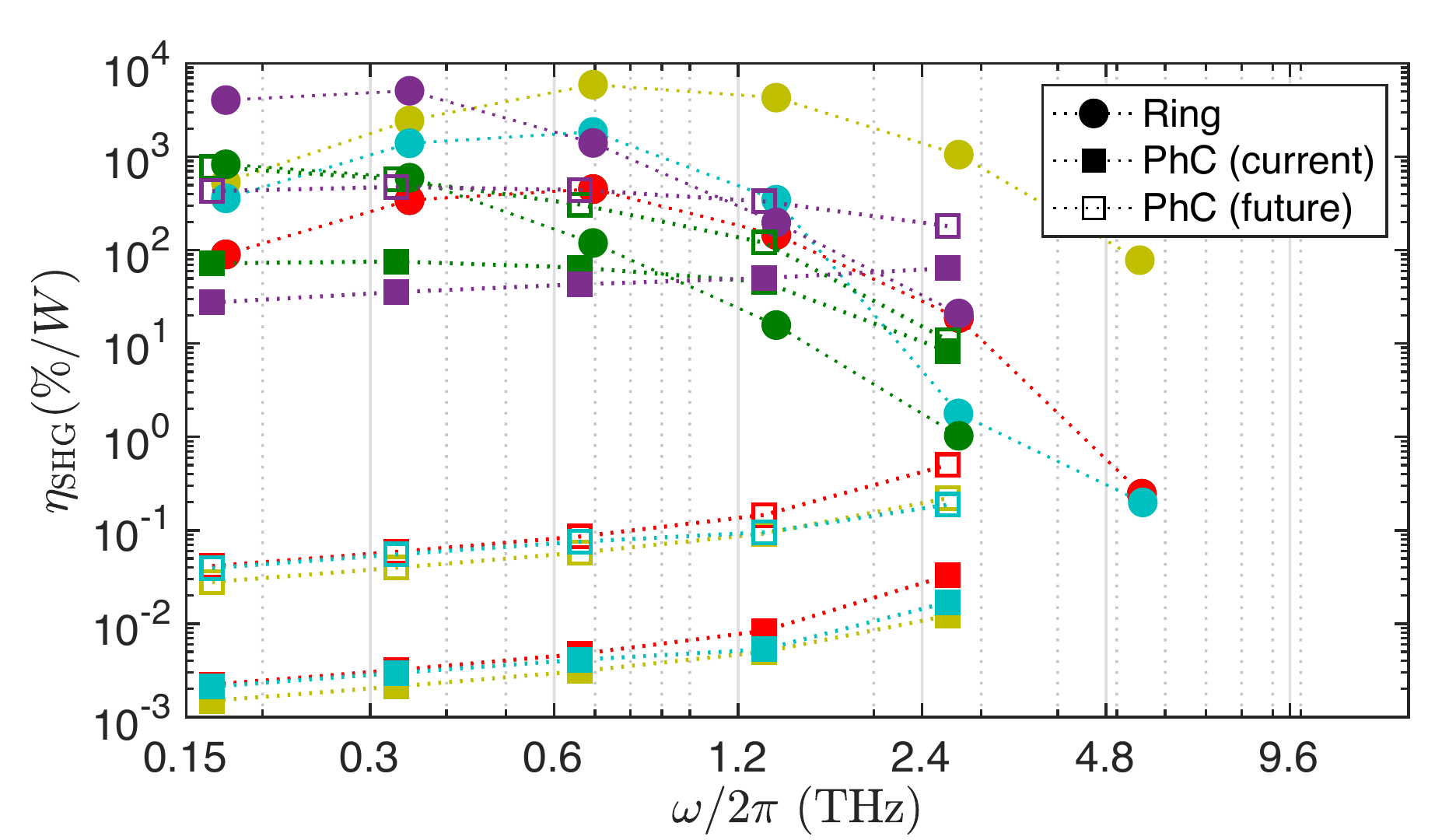} % 08.05.21
    \caption{Maximum second harmonic generation efficiency for doubling frequencies, in \%/W. Colors denote the materials as in~\figref{thz-params}. Solid and hollow markers represent the current design and near-future designs with improved $Q$, respectively. For the photonic crystal cavities, $Q^\text{r}_a = 358,000$ ($10^6$) and $Q^\text{r}_b = 41,600$ ($10^5$) for the current (near-future) designs. The cross-sectional area of the nonlinear material is 1 $\mu$m $\times$ 1$\mu$m for 331 GHz, and inversely proportional to the square of frequencies. %For the ring cavities, we assume $Q^\text{r}_{a,b} = 10^6$, and the cross-sectional area of the nonlinear material is fixed to 100 nm $\times$ 100 nm at every frequency.
    For the ring cavities, we calculate $Q^\text{r}_{a,b} \geq 10^6$, and the cross-sectional area of the nonlinear material is fixed to 100 nm $\times$ 100 nm at every frequency. For all calculations, the silicon is assumed to be lossless.}
    \label{fig:efficiency}
\end{figure}

\figref{efficiency} plots the THz SHG conversion efficiencies in the nondepleted pump regime for different nonlinear materials. For the PhC design, we also show projected efficiencies for improved radiation quality factor $Q_{a(b)}^{r}$. We designed PhC cavities with a nonlinear material volume small enough for $Q_{a(b)}$ to be limited by radiation loss, and ring cavities with $Q_{a(b)}$ limited largely by material loss. For simplicity, we ignore the index dispersion, considering instead the dispersion of loss and nonlinear coefficients. 

The conversion efficiencies are above $10^3 \%$/W for the current designs across most of the THz gap, indicating near-unity conversion efficiencies with input powers on the order of hundreds of milliwatts. Analysis of the absolute conversion efficiency (depleted pump regime) provides the output power at each stage of the cascaded process, shown in~\figref{cavity}(a) for the ring~\cite{suppConversionEff}. %(see further discussion in Supplementary Section 1). 

% \FloatBarrier

%%%%%%%%%%%%%%%%%%%%%%%%%%%%%%%%%%%%%%%%%%%%%%%%%%%%%%%%%%%%%%%%%%%%%%%%%%%%%%%%%%%%%%%%%%%%%%%%%
% DISCUSSION
%%%%%%%%%%%%%%%%%%%%%%%%%%%%%%%%%%%%%%%%%%%%%%%%%%%%%%%%%%%%%%%%%%%%%%%%%%%%%%%%%%%%%%%%%%%%%%%%%

\section{Discussion}
We theoretically estimated the first- and second-order susceptibility of several promising materials in the terahertz region: GaAs, GaP, ZnTe, LiNbO$_3$, and LiTaO$_3$. Quantum materials such as charge density wave materials (TaS$_2$), excitonic insulators (Ta$_2$NiS$_5$), and collective excitations in superconductors such as niobium nitride may enable even higher second harmonic efficiency or signal amplification, though we leave a detailed analysis with proper accounting of loss for future work. As the ultimate limit to field concentration into a nonlinear material, future work should also consider THz-band nonlinearities in graphene~\cite{hafez2020terahertz}.

A key to minimizing optical loss is to embed such materials in regions of high field concentration: hybrid dielectric cavities employing dielectric tip structures in photonic crystal cavities and ring resonators. We anticipate that substantial efficiency gains will be possible by co-design of materials, resonators, and SHG phase matching. These efforts also call for experimental studies of THz-spectrum nonlinearities, cavities, loss mechanisms, etc. These dielectric field concentration structures are also promising as low-loss alternatives to doubly resonant metallic nanoantennas, which underpin an exciting recent proposal for THz-to-optical parametric frequency conversion at the single photon level~\cite{roelli2020molecular}.

From the calculations of non-depleted and absolute conversion efficiencies of our devices over a sequence of cascaded frequency doubling steps, we estimate the opportunity to fill the THz gap with 1 W of input power, generating THz radiation with substantially higher efficiency than electronic sources alone, but without the need for cryogenic cooling in THz lasers. %To handle the high input powers assumed in~\figref{cavity}, larger cavities or coherent combination of multiple cavity outputs between stages may be used to operate below damage threshold  
It is clear from the curves in~\figref{cavity} that the fall-off in output power is very sudden, since cascaded losses compound in each SHG step. To sustain high output power after many SHG steps, we therefore considered a rather high input power of up to 4 W at the $\sim$100 GHz seed frequency. Due to limitations imposed by dielectric breakdown, high input powers may require larger cavities in the early stages of frequency conversion to distribute gain, or coherent combination of multiple cavity outputs between stages to operate below damage threshold while compensating losses in each step~\cite{suppDielecBreakdown}. %(See Supplementary Section 4 for an analysis of dielectric breakdown.)

%%%%%%%%%%%%%%%%%%%%%%%%%%%%%%%%%%%%%%%%%%%%%%%%%%%%%%%%%%%%%%%%%%%%%%%%%%%%%%%%%%%%%%%%%%%%%%%%%
% CONCLUSION
%%%%%%%%%%%%%%%%%%%%%%%%%%%%%%%%%%%%%%%%%%%%%%%%%%%%%%%%%%%%%%%%%%%%%%%%%%%%%%%%%%%%%%%%%%%%%%%%%

\section{Conclusion}
We proposed a new approach for light sources in the THz spectrum based on cascaded second harmonic generation pumped by low-noise electronic oscillators. As opposed to laser sources, this approach requires no population inversion; and in contrast to electronic sources, where ohmic loss limits high-frequency operation, our dominant loss originates from dielectric absorption and radiation. It leverages high quality factor dielectric resonators with phonon-resonance-enhanced second-order nonlinearity for parametric frequency conversion, which is inherently power-preserving. In particular, our modeling shows that the proposed devices provide sufficiently high SHG conversion efficiency to be cascaded over multiple octaves, provided sufficiently high input power from the electronic source. Moreover, sum/difference frequency generation (SFG/DFG) would enable arbitrary frequencies. Using a combination of cascaded SHG, SFG, and DFG, our approach opens the door to compact, low-cost, and room temperature devices that deliver high power THz radiation at any frequencies in the THz gap, and that may be extended into the mid-IR and beyond.  

%%%%%%%%%%%%%%%%%%%%%%%%%%%%%%%%%%%%%%%%%%%%%%%%%%%%%%%%%%%%%%%%%%%%%%%%%%%%%%%%%%%%%%%%%%%%%%%%%
% END MATTER
%%%%%%%%%%%%%%%%%%%%%%%%%%%%%%%%%%%%%%%%%%%%%%%%%%%%%%%%%%%%%%%%%%%%%%%%%%%%%%%%%%%%%%%%%%%%%%%%%

\section*{Acknowledgments}
The authors thank C. Panuski and M. El Kabbash for helpful feedback. This work was supported in part by the Defense Advanced Research Projects Agency (DARPA) DRINQS (HR001118S0024) program. L.A. was supported by the National Science Foundation Graduate Research Fellowship (Grant No. 1745302). H.C. was supported in part by the Claude E. Shannon Research Assistantship and Samsung Scholarship.

\section*{Disclosures}
The authors declare no conflicts of interest.

\section*{Contributions}
D.E. and H.C. conceived and initiated the presented idea. L.A. analyzed nonlinear susceptibilities. H.C. proposed anisotropic anharmonicity for nonlinearity correction. H.C. designed the doubly resonant PhC cavity. L.A. designed the ring cavity. M.H. simulated equation of motion in non-depletion regime. All authors contributed in writing manuscript.

\bibliography{biblio}

\end{document}

\preprint{APS/123-QED}

%%%%%%%%%%%%%%%%%%%%%%%%%%%%%%%%%%%%%%%%%%%%%%%%%%%%%%%%%%%%%%%%%%%%%%%%%%%%%%%%%%%%%%%%%%%%%%%%%
% TITLE AND AUTHOR INFO
%%%%%%%%%%%%%%%%%%%%%%%%%%%%%%%%%%%%%%%%%%%%%%%%%%%%%%%%%%%%%%%%%%%%%%%%%%%%%%%%%%%%%%%%%%%%%%%%%

\title{Supplemental Information: Terahertz light sources by electronic-oscillator-driven \\ second harmonic generation in extreme-confinement cavities}

\author{Lamia Ateshian}
\altaffiliation{These authors contributed equally to this work.}
\author{Hyeongrak Choi}
\altaffiliation{These authors contributed equally to this work.}
\author{Mikkel Heuck}
\author{Dirk Englund}
\email{englund@mit.edu}
\affiliation{%
 Research Laboratory of Electronics, Massachusetts Institute of Technology, Cambridge, Massachusetts 02139, USA
}%
\affiliation{%
 Department of Electrical Engineering and Computer Science, Massachusetts Institute of Technology, Cambridge, Massachusetts 02139, USA
}%

\begin{abstract}
\end{abstract}

\maketitle

% Supplemental content
\onecolumngrid
% \input{supp-content}

%%%%%%%%%%%%%%%%%%%%%%%%%%%%%%%%%%%%%%%%%%%%%%%%%%%%%%%%%%%%%%%%%%%%%%%%%%%%%%%%%%%%%%%%%%%%%%%%%
% CONVERSION EFFICIENCY
%%%%%%%%%%%%%%%%%%%%%%%%%%%%%%%%%%%%%%%%%%%%%%%%%%%%%%%%%%%%%%%%%%%%%%%%%%%%%%%%%%%%%%%%%%%%%%%%%

\section{Conversion Efficiency} \applab{conversion efficiency}
The conversion efficiency of second-harmonic generation (SHG) can be derived from the Hamiltonian describing a doubly-resonant cavity coupled to waveguide modes at both fundamental and second-harmonic (SH) frequency~\cite{guo2016second}
\begin{align}\eqlab{Hamiltonian}
    \mathcal{H}/\hbar = \delta_a \hat{a}^\dagger \hat{a} + \delta_b \hat{b}^\dagger \hat{b} +  \left[ g(\hat{a}^\dagger)^2 \hat{b} + g^*\hat{a}^2\hat{b}^\dagger \right] + \sqrt{2\kappa_a^c}(\hat{\epsilon}_p^\dagger \hat{a}+\hat{\epsilon}_p \hat{a}^\dagger) + \sqrt{2\kappa_b^c}(\hat{\epsilon}_{\rm{SHG}}^\dagger \hat{b}+\hat{\epsilon}_{\rm{SHG}} \hat{b}^\dagger).
\end{align}
The operators $\hat{a}$ and $\hat{b}$ annihilate a photon from the FD and SH mode of the cavity, while $\hat{\epsilon}_p$ ($\hat{\epsilon}_{\rm{SHG}}$) is the annihilation operator for input (output) photons in the waveguide. The resonance frequencies of the cavity modes are $\omega_a$ and $\omega_b=2\omega_a$ while the carrier frequency of the input (output) field is $\omega_p$ ($2\omega_p)$. Note that~\eqref{Hamiltonian} is expressed in a frame rotating with $\omega_p$ and $2\omega_p$ for $\hat{a}$ and $\hat{b}$, which introduces the detunings $\delta_a = \omega_a - \omega_p$ and $\delta_b = \omega_b - 2\omega_p$. The cavity-waveguide coupling rate of the fundamental (SH) mode is $\kappa_a^c$ ($\kappa_b^c)$ and the nonlinear coupling rate $g$ can be derived from perturbation theory~\cite{choi2017self, joannopoulos2008molding}
% % 
% \begin{align}\eqlab{g def app}
%     g = \sqrt{\frac{\hbar\omega_a^2 \omega_b}{\epsilon_0}}\cdot\frac{\int dV \chi^{(2)}(\bm{r}) u_a^2(\bm{r})u^*_b(\bm{r})}{\int dV \epsilon(\bm{r})|u_a(\bm{r})|^2\sqrt{\int dV \epsilon(\bm{r})|u
%     _b(\bm{r})|^2}}.
% \end{align}
% % 
% 
\begin{align}\eqlab{g def app}
    g &= \chi^{(2)}_\text{eff}\sqrt{\frac{\hbar\omega_a^2\omega_b}{\epsilon_0}}\frac{\tilde{\beta}}{\sqrt{(\lambda_a/n_a)^3}},
\end{align}
where the normalized field overlap factor, $\tilde{\beta}$, is given by
\begin{align}\label{eq:beta}
    \tilde{\beta} &= \frac{1}{\chi^{(2)}_\text{eff}} \frac{\int dV \chi_{ijk}^{(2)}(\vec{r})  E_{a,i}(\vec{r})E_{a,j}(\vec{r})E^*_{b,k}(\vec{r})}{\int dV \epsilon(\vec{r})\left|\vec{E}_a(\vec{r})\right|^2\sqrt{\int dV \epsilon(\vec{r})\left|\vec{E}_b(\vec{r})\right|^2}}  \sqrt{\frac{\lambda_a^3}{n_a^3}}.
\end{align}
Note that $\tilde{\beta}$ is normalized to the wavelength in the nonlinear medium in contrast to $\bar{\beta}$ in \cite{Lin2016}, which is normalized to the vacuum wavelength. The normalization is chosen for fair comparison between nonlinear materials. We used Einstein notation implying summation over the repeated Cartesian subscripts $(i,j,k)\in (x,y,z)$. The field profiles of the unperturbed cavity modes are $\vec{E}_a(\vec{r})$ and $\vec{E}_b(\vec{r})$ and $n_a$ is the refractive index at $\omega_a$. The equations of motion of the cavity modes $\hat{a}$ and $\hat{b}$ can be derived from the Heisenberg equation of motion $i\hbar\frac{d\hat{O}}{dt} = \left[\hat{O}, \mathcal{H}\right]$ where $\hat{O}$ is the cavity mode operators
\begin{subequations}\eqlab{eoms quantum}
\begin{align}
    \frac{d\hat{a}}{dt} &= [-i\delta_a-\kappa_a^c]\hat{a}-i2g\hat{a}^\dagger\hat{b}+ i\sqrt{2\kappa_a^c}\hat{\epsilon}_p, \\
    \frac{d\hat{b}}{dt} &= [-i\delta_b-\kappa_b^c]\hat{b}-ig^*\hat{a}^2.
\end{align}
\end{subequations}
% 
% Note that our kappa is not "intensity" decay rate, but "Field" decay rate, so kappa = omega/2Q.
Note that~\eqref{eoms quantum} assumes that there is no incident field at $\omega_b$. Assuming that all fields are in strong coherent states, we make the replacement of operators with c-numbers~\cite{Vernon2016} $\hat{a}\rightarrow \alpha_a$,  $\hat{b}\rightarrow \alpha_b$, $\hat{\epsilon}_p \rightarrow \epsilon_p$, $\hat{\epsilon}_{\rm{SHG}} \rightarrow \epsilon_{\rm{SHG}}$. This results in equations of motion and input-output relations for the c-numbers that are normalized such that the square of their absolute values correspond to the number of photons in the cavity modes and photon flux in the waveguide
\begin{subequations}\eqlab{eoms classical}
\begin{align}
    \frac{d\alpha_a}{dt} &= [-i\delta_a-\kappa_a]\alpha_a-i2g\alpha_a^*\alpha_b+ i\sqrt{2\kappa_a^c}\epsilon_p, \eqlab{eoms classical a}\\
    \frac{d\alpha_b}{dt} &= [-i\delta_b-\kappa_b]\alpha_b-ig^*\alpha_a^2, \eqlab{eoms classical b}\\
    \epsilon_p^{\rm{out}} &= \epsilon_p - \sqrt{2\kappa_a^c}\alpha_a, \eqlab{eoms classical eps a}\\
    \epsilon_{\rm{SHG}} &= - \sqrt{2\kappa_b^c}\alpha_b. \eqlab{eoms classical eps b}
\end{align}
\end{subequations}
In \eqref{eoms classical}, we introduced the total decay rates, $\kappa_{a(b)} = \kappa_{a(b)}^c + \kappa_{a(b)}^m + \kappa_{a(b)}^r$, where the superscripts correspond to waveguide-coupling ($c$), material absorption ($m$), and radiation loss ($r$). We consider a continuous wave (CW) input and steady-state, where $d\alpha_{a(b)}/dt = 0$. In the limit of low conversion efficiency, we use the non-depleted pump approximation, which neglects the nonlinear term in~\eqref{eoms classical a}. In this case, we have  
% The approximation has two consequences; 1) fundamental mode maintains classical coherent state, and the annihilation operator can be replaced by c-number we call $\alpha$; 2) zeroth order calculation of $\alpha$ ignores the coupling with the second-harmonic mode.
% 
\begin{align}\eqlab{eoms sol SS NP}
    \alpha_a  = \frac{i\sqrt{2\kappa_{a}^c}\epsilon_p}{-i\delta_a-\kappa_a}, ~~\text{and} ~~ \alpha_b = \frac{ig^*}{-i\delta_b-\kappa_b}\alpha_a^2.
\end{align}
The optical power in the input and output waveguides are $P_p = \hbar\omega_a|\epsilon_{p}|^2$ and $P_{\rm{SHG}} = \hbar\omega_b|\epsilon_{\rm{SHG}}|^2$, respectively. From~\eqsref{eoms sol SS NP}{eoms classical eps b}, we have
\begin{align}\eqlab{PSHG a}
    P_{\rm{SHG}} &= \hbar\omega_b \big|\sqrt{2\kappa_{b}^c}\alpha_b\big|^2  = \hbar\omega_b 2|g|^2 \frac{\kappa_{b}^c}{\delta_b^2 + \kappa_b^2} |\alpha_a|^4 =  \frac{\hbar\omega_b 2|g|^2 \kappa_{b}^c}{\delta_b^2 + \kappa_b^2} \frac{4(\kappa_{a}^c)^2}{\big(\delta_a^2+\kappa_a^2\big)^2} |\epsilon|^4 =  \frac{\hbar\omega_b 2|g|^2 \kappa_{b}^c}{\delta_b^2 + \kappa_b^2} \frac{4(\kappa_{a}^c)^2}{\big(\delta_a^2+\kappa_a^2\big)^2} \frac{P_p^2}{\hbar^2\omega_a^2}. 
    % P_{\rm{out}} &= \hbar\omega_b |\epsilon_{\rm{out}}|^2 = 2\hbar\omega_a 8 |g|^2 \frac{\kappa_{b}^c}{\delta_b^2 + \kappa_b^2} \frac{\kappa_{ac}^2}{\big(\delta_a^2+\kappa_a^2\big)^2} |\epsilon|^4.
\end{align}   
The output power is maximized when $\delta_a=\delta_b=0$, in which case~\eqref{PSHG a} becomes
\begin{align}\eqlab{PSHG b}
    P_{\rm{SHG}} &= \frac{16}{\hbar\omega_a} |g|^2 \frac{2Q_b}{\omega_b} \frac{4Q_a^2}{\omega_a^2} \left(\frac{\kappa_{a}^c}{\kappa_a}\right)^{\!\!2}\frac{\kappa_{b}^c }{\kappa_b} P_{p}^2 = \frac{64}{\hbar\omega_a^4} |g|^2 Q_a^2 Q_b  \left(\frac{Q_a}{Q_{a}^c}\right)^{\!\!2} \frac{Q_b }{Q_{b}^c}  P_{p}^2,
\end{align}   
where we used $Q_{a(b)} =\omega_{a(b)}/(2\kappa_{a(b)})$. The non-depleted conversion efficiency is therefore given by
\begin{align}  \eqlab{eta SHG}
    \eta_{\rm{SHG}} \equiv \frac{P_{\rm{SHG}}}{P_{p}^2} &= \frac{64}{\hbar\omega_a^4} |g|^2 Q_a^2 Q_b \eta_c,
\end{align}
where the input-output coupling efficiency is defined as
\begin{align}  \eqlab{eta c}
    \eta_{c} \equiv \left(\frac{Q_a}{Q_{a}^c}\right)^{\!\!2} \frac{Q_b }{Q_{b}^c}.
\end{align}
From~\eqref{PSHG a} we find that the non-depleted conversion efficiency is maximized when $\kappa_{a(b)}^c = \kappa_{a(b)}^m + \kappa_{a(b)}^r$ corresponding to critical coupling. \\

In the limit of large conversion efficiency, we solve~\eqref{eoms classical} to evaluate the absolute conversion efficiency, $\eta_{\rm{SHG}}^{\rm{abs}}\equiv P_{\rm{SHG}}/P_p$, where $P_{\rm{SHG}}$ is calculated without making the non-depleted pump approximation. In this case, the optimum coupling $Q$ must be determined via numerical optimization. In~\figref{eta abs example}(a) we show an example of the absolute conversion efficiency as a function of input power and the corresponding values of $Q_{a(b)}^c$ are plotted in~\figref{eta abs example}(b).\\
\begin{figure}[htbp!]
    \centering
    \includegraphics[width=0.35\textwidth]{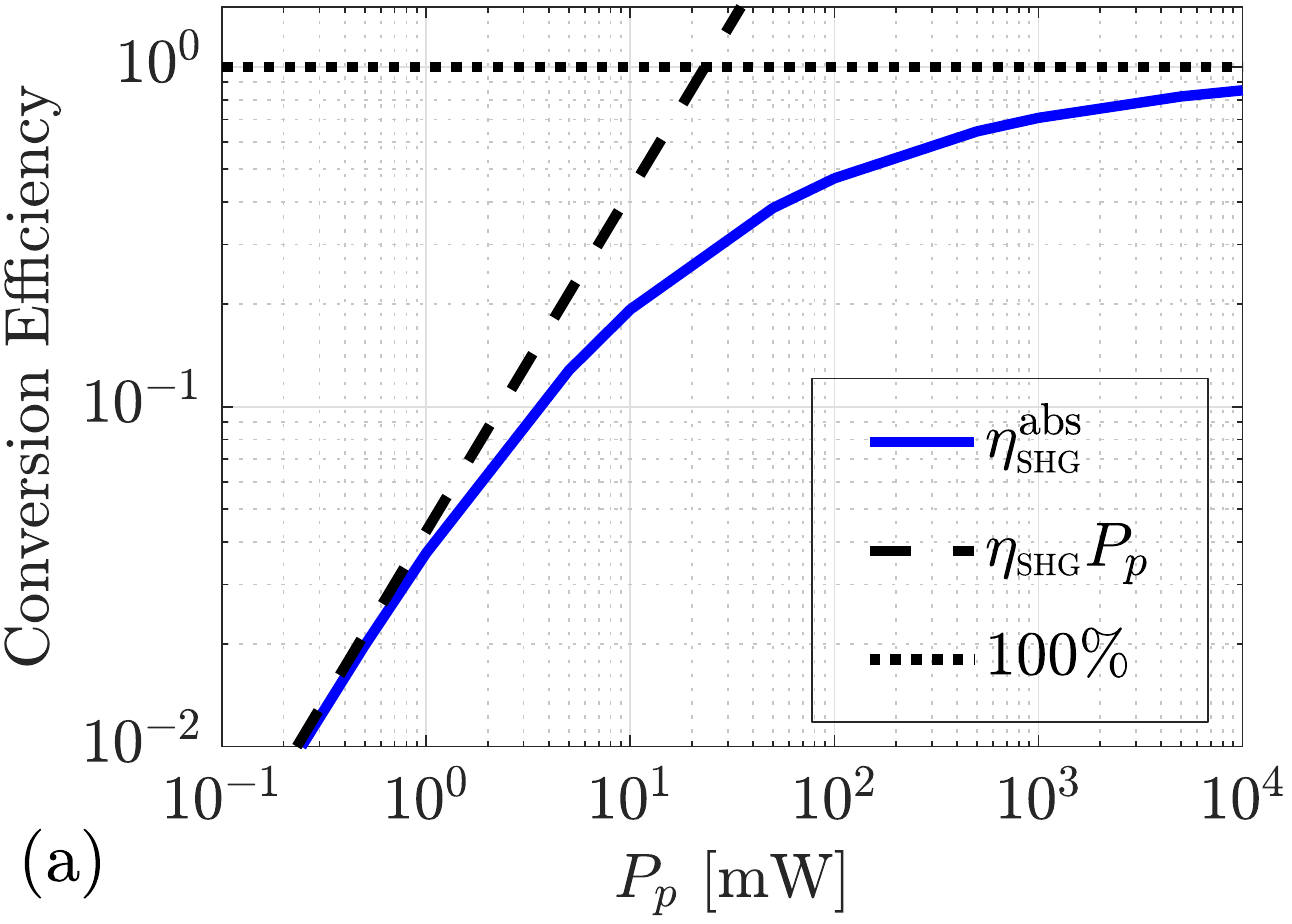}
    \hspace{8mm}
    \includegraphics[width=0.35\textwidth]{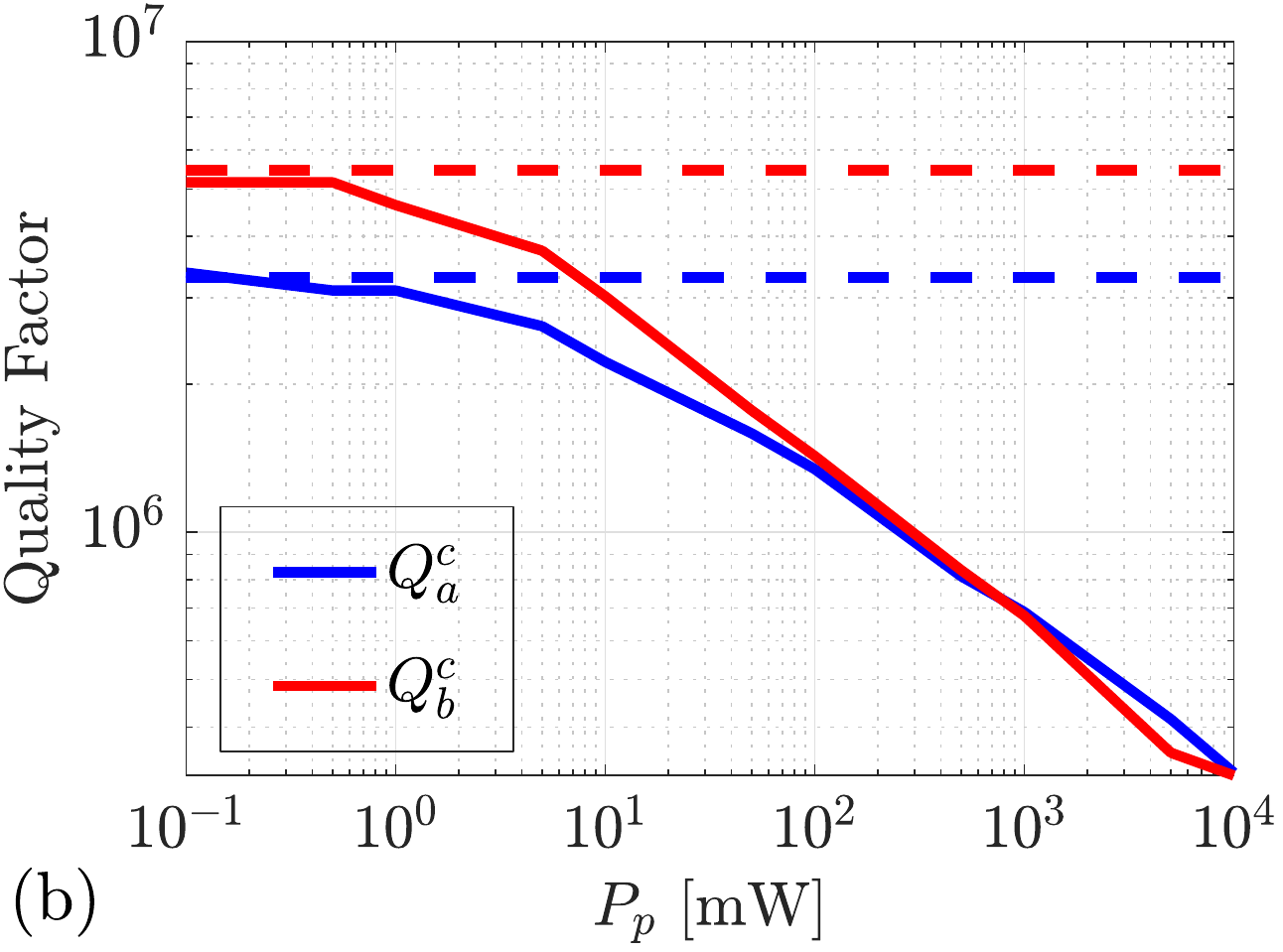}
    \caption{Example of a calculation of the absolute conversion efficiency for the ring cavity design. (a) $\eta_{\rm{SHG}}^{\rm{abs}}$ is plotted as a function of input power. The black dashed line plots $\eta_{\rm{SHG}}P_p$ from~\eqref{eta SHG}, which asymptotically approaches $\eta_{\rm{SHG}}^{\rm{abs}}$ at low input power. (b) Optimum choice of coupling quality factors corresponding to $\eta_{\rm{SHG}}^{\rm{abs}}$ in (a). $Q_{a(b)}^c$ approach their values at critical coupling (dashed lines) at low power levels as expected. The parameters used in~\eqref{eoms classical} to calculate $\eta_{\rm{SHG}}^{\rm{abs}}$ correspond to a ring cavity design using GaP as the nonlinear material at a frequency of $\omega_a/2\pi = 1385\,$GHz. The parameter values are: $Q_a^r= 6.0\!\times\!10^{6}$, $Q_b^r = 2.0\!\times\!10^{13}, $$Q_a^m = 7.3\!\times\!10^{6}$, $Q_b^m = 5.5\!\times\!10^{6}$,   $|\tilde{\beta}| = 2.4\!\times\!10^{-5}$, $n_a = 3.32$, and $\chi_{\rm{eff}}^{(2)} = 1.2\!\times\!10^{-10}$m/V.} 
    \figlab{eta abs example}
\end{figure}

To calculate the power at each frequency of the cascaded SHG plotted in
% ~\figref{cavity}, 
Fig. 1 of the main text,
we first calculate $\eta_{\rm{SHG}}^{\rm{abs}}$ as a function of input power at each frequency for the different materials illustrated in
% ~\figref{efficiency}.
Fig. 4.
At each frequency, the material resulting in the largest conversion efficiency is chosen and the output power at $\omega_b$ becomes the input power at the next stage of the cascade.~\figsref{phc eta abs}{ring eta abs} plot the power available at each frequency in the cascade process for three different values of the starting power, $P_{\rm{in}}$, for our PhC and ring cavity designs. The figures also plot the corresponding intra-cavity energy in the fundamental and second-harmonic modes. The parameters corresponding to optimum performance are listed in Table~\ref{tb: phc parameters} for the PhC designs and Table~\ref{tb: ring parameters} for ring resonators (corresponding to the results in~\figref{ring eta abs}).
\begin{figure}[htbp!]
    \centering
    \includegraphics[width=0.35\textwidth]{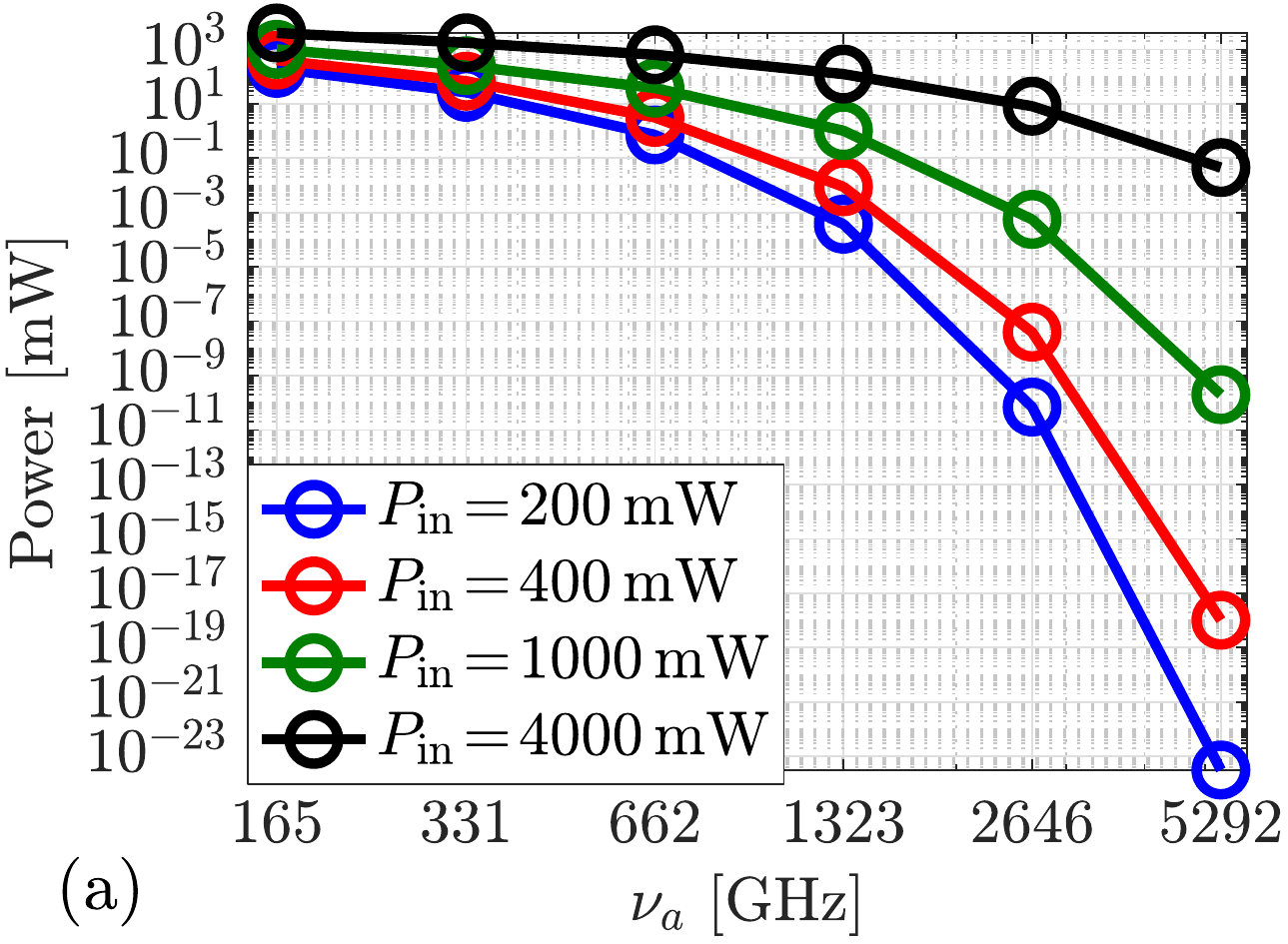}
    \hspace{8mm}
    \includegraphics[width=0.35\textwidth]{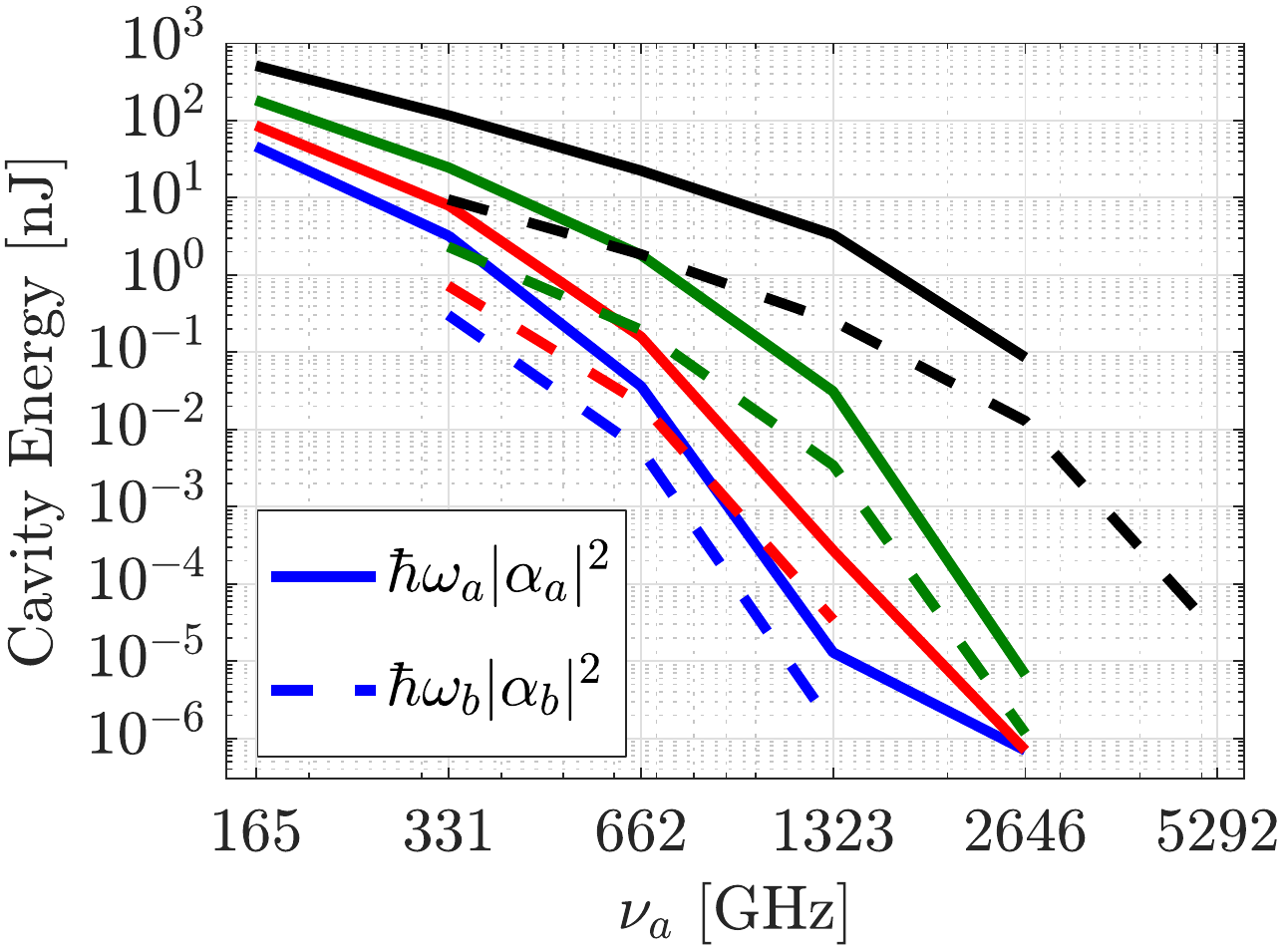}
    \caption{Example of cascading using the best possible PhC design at each frequency. (a) Output power as a function of frequency for different input power. (b) Intra-cavity energy corresponding to the output powers in (a). Solid and dashed lines plot the energy in the fundamental and SH modes, respectively. Simulation parameters are listed in Table~\ref{tb: phc parameters}.} 
    \figlab{phc eta abs}
\end{figure}
The listed material properties ($\chi_{\rm{eff}}^{(2)}$ and $n_a$) are found in~\appref{nonlinear material analysis} and the cavity parameters ($Q_{a(b)}^r$, $Q_{a(b)}^m$, and $\tilde{\beta}$) are found in~\appref{app cavity designs}. The coupling quality factors, $Q_{a(b)}^c$, are found by maximizing the absolute conversion efficiency as described above. 

\subsection{Radiation loss by SH polarization} \applab{SH rad loss}
We estimate the conversion efficiency from the coupling of an induced SH polarization to the SH cavity mode. Analogous to spontaneous emission of an atom in cavity-QED, the SH polarization can directly radiate into free-space. To the best of our knowledge, this direct SH radiation has never been studied in the literature. We neglected this loss mechanism in our estimation and leave a quantitative analysis for future studies, but the discussion below briefly outlines how it should be incorporated.

The SH polarization field is driven by a source with a spatial distribution determined by $\vec{E}_a^2(\vec{r})$ and its amplitude is proportional to $\alpha_a^2$. From the equation of motion of the fundamental field amplitude, $\alpha_a$, Eq.~(S5a), we know that the rate of change of $\alpha_a$ originating from its coupling to modes at $2\omega_a$ is proportional to $g\alpha_a^*\alpha_{m}$, where $\alpha_{m}$ is the amplitude of the $m$th (leaky) mode at $2\omega_a$. Since the source of these modes is the SH polarization field, we have $\alpha_m\propto \alpha_a^2$, which results in a loss term in Eq.~(S5a) proportional to $|\alpha_a|^2$ 
% Direct SH radiation modifies the equation of motion of the fundamental field, $\alpha_a$ (Eq.~S5a), as;
\begin{align}
    \frac{d\alpha_a}{dt} &= [-i\delta_a-\kappa_a-\kappa_m|\alpha_a|^2]\alpha_a-i2g\alpha_a^*\alpha_b+ i\sqrt{2\kappa_a^c}\epsilon_p,
\end{align}
% where $\kappa_n|\alpha_a|^2\alpha_a$ is added.
where the proportionality factor is denoted by $\kappa_m$. This term is analogous to two-photon absorption of the nonlinear medium. Due to the technical subtlety of determining $\kappa_m$, we leave it for future research. Nevertheless, this additional term can be neglected in the non-depleted regime. Thus, all the conversion efficiencies in \%/W are unaffected by this term. 
\begin{table}[htbp!]
\begin{center}
% \begin{adjustwidth}{-0.5in}{-0.5in}
\begin{tabular}{| c | c | c | c | c | c | c | c | c | c | c |}
\hline
 \textbf{Freq. [GHz]} & \textbf{Material}  & \bm{$\chi_{\rm{eff}}^{(2)}$} \textbf{[m/V]} & \bm{$\tilde{\beta}$} & \bm{$Q_a^r$} & \bm{$Q_a^m$} & \bm{$Q_a^c$} & \bm{$Q_b^r$} & \bm{$Q_b^m$} & \bm{$Q_b^c$} & \bm{$n_a$}  \\ 
 \hline
 165.4 & LiTaO$_3$   & 2.46$\times10^{-8}$ & 6.26$\times10^{-6}$ & 3.58$\times10^{5}$ & 2.06$\times10^{6}$ & 1.8$\times10^{5}$  & 4.16$\times10^{4}$ & 6.14$\times10^{5}$ & 1.9$\times10^{4}$ & 6.13\\ 
 \hline
 330.7 & LiTaO$_3$   & 2.49$\times10^{-8}$ & 3.72$\times10^{-6}$ & 3.58$\times10^{5}$ & 1.05$\times10^{6}$ & 2.3$\times10^{5}$  & 4.16$\times10^{4}$ & 3.10$\times10^{5}$ & 2.5$\times10^{4}$ & 6.14\\ 
 \hline
 661.5 & LiTaO$_3$   & 2.62$\times10^{-8}$ & 2.20$\times10^{-6}$ & 3.58$\times10^{5}$ & 5.16$\times10^{5}$ & 2.0$\times10^{5}$  & 4.16$\times10^{4}$ & 1.42$\times10^{5}$ & 2.9$\times10^{4}$ & 6.17 \\ 
 \hline
 1323 & LiNbO$_3$   & 1.55$\times10^{-8}$ & 1.71$\times10^{-6}$ & 3.58$\times10^{5}$ & 8.84$\times10^{5}$ & 2.7$\times10^{5}$  & 4.16$\times10^{4}$ & 2.12$\times10^{5}$ & 3.5$\times10^{4}$ & 5.16\\ 
 \hline
 2646 & LiNbO$_3$   & 3.02$\times10^{-8}$ & 9.56$\times10^{-7}$ & 3.58$\times10^{5}$ & 3.53$\times10^{5}$ & 1.9$\times10^{5}$  & 4.16$\times10^{4}$ & 2.85$\times10^{4}$ & 1.7$\times10^{4}$ & 5.37 \\ 
 \hline
\end{tabular}
% \end{adjustwidth}
\end{center}
\caption{Parameters used to calculate the cascaded output power for the PhC cavity design in~\figref{phc eta abs} using a starting input power of $P_{\rm{in}}= 1000\,$mW.}
\label{tb: phc parameters}
\end{table}
% % 
% \begin{figure}[h!]
%     \centering
%     \includegraphics[width=0.3\textwidth]{figures/z_phc_Emax_Silicon_cascade.pdf}
%     \hspace{4mm}
%     \includegraphics[width=0.3\textwidth]{figures/z_phc_Emax_Air_cascade.pdf}
%     \hspace{4mm}
%     \includegraphics[width=0.3\textwidth]{figures/z_phc_Emax_Nonlinear Material_cascade.pdf}
%     \caption{Maximum field strength in each of the cavity materials corresponding to the power and intra-cavity energy in~\figref{phc eta abs}.} 
%     \figlab{phc Emax}
% \end{figure}
% % 

% 
\begin{figure}[htbp!]
    \centering
    \includegraphics[width=0.35\textwidth]{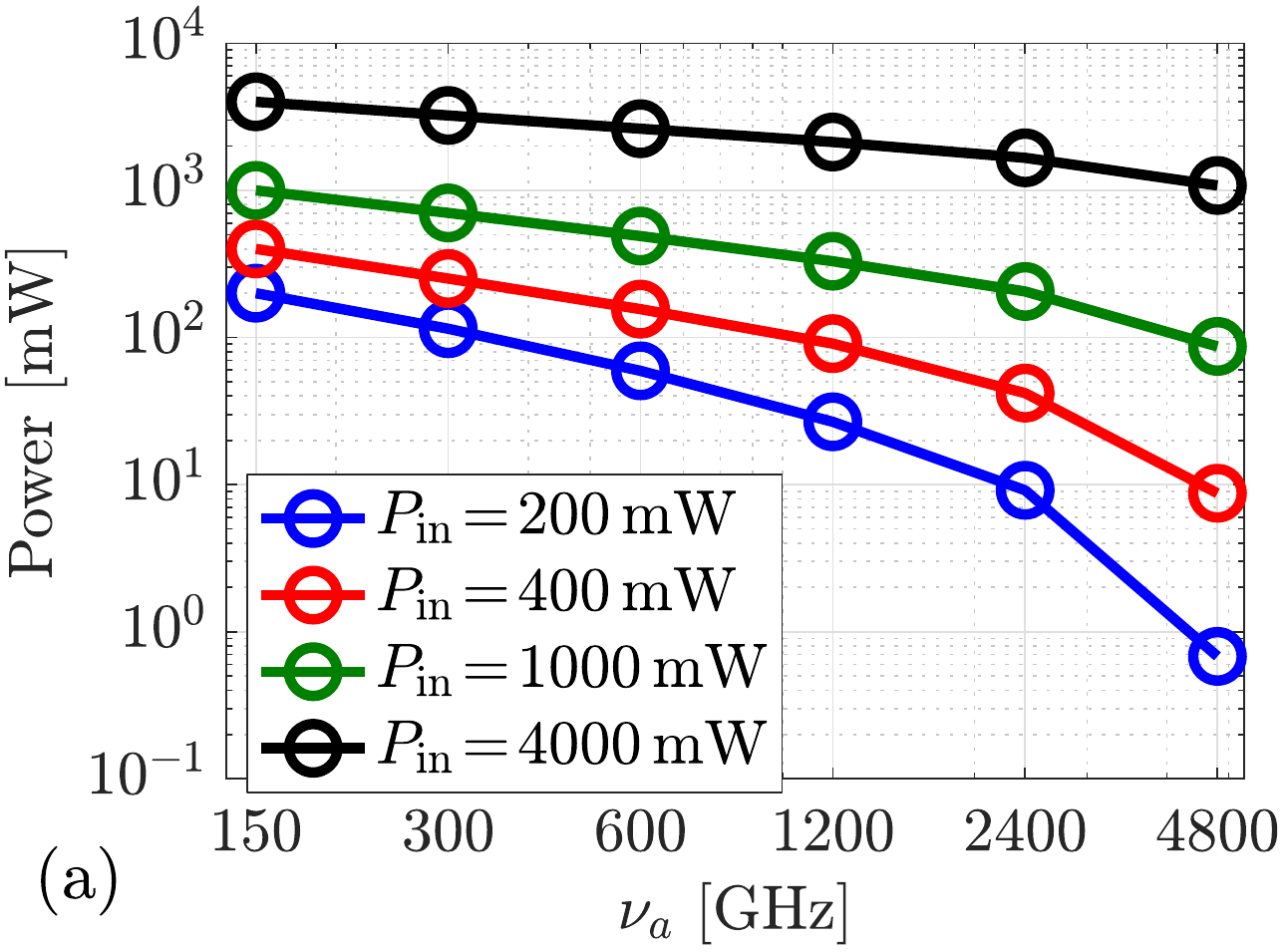}
    \hspace{8mm}
    \includegraphics[width=0.35\textwidth]{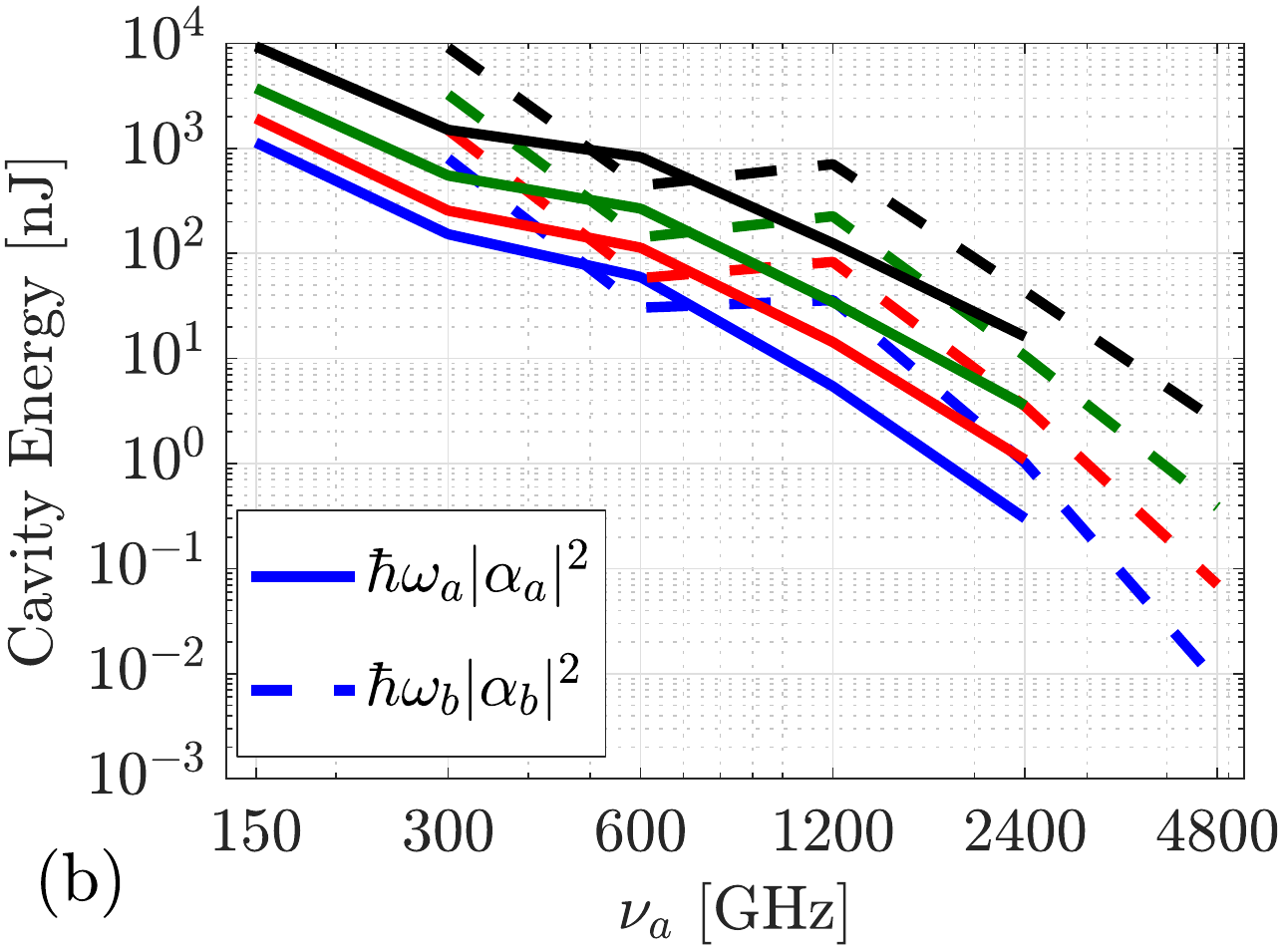}
    \caption{Example of cascading using the best possible ring design at each frequency. (a) Output power as a function of frequency for different input power. (b) Intra-cavity energy corresponding to the output powers in (a). Solid and dashed lines plot the energy in the fundamental and SH modes, respectively. Simulation parameters are listed in Table~\ref{tb: ring parameters}.} 
    \figlab{ring eta abs}
\end{figure}

\begin{table}[htbp!]
\begin{center}
% \begin{adjustwidth}{-0.5in}{-0.5in}
\begin{tabular}{| c | c | c | c | c | c | c | c | c | c | c |}
\hline
 \textbf{Freq. [GHz]} & \textbf{Material}  & \bm{$\chi_{\rm{eff}}^{(2)}$} \textbf{[m/V]} & \bm{$\tilde{\beta}$} & \bm{$Q_a^r$} & \bm{$Q_a^m$} & \bm{$Q_a^c$} & \bm{$Q_b^r$} & \bm{$Q_b^m$} & \bm{$Q_b^c$} & \bm{$n_a$}\\ 
 \hline
 175 & LiNbO$_3$   & 1.3$\times10^{-8}$ & 3.7$\times10^{-8}$ & 2.4$\times10^{7}$ & 1.1$\times10^{8}$ & 4.0$\times10^{6}$  & 2.7$\times10^{14}$ & 8.2$\times10^{7}$ & 1.0$\times10^{7}$ & 5.10\\ 
 \hline
 350 & LiNbO$_3$   & 1.3$\times10^{-8}$ & 1.4$\times10^{-7}$ & 1.9$\times10^{7}$ & 1.3$\times10^{7}$ & 1.7$\times10^{6}$  & 1.8$\times10^{14}$ & 1.0$\times10^{7}$ & 1.3$\times10^{6}$ & 5.10\\ 
 \hline
 700 & GaP         & 1.1$\times10^{-10}$ & 6.8$\times10^{-6}$ & 1.3$\times10^{7}$ & 5.5$\times10^{7}$ & 2.4$\times10^{6}$   & 8.2$\times10^{13}$ & 4.3$\times10^{7}$ & 6.1$\times10^{6}$ & 3.32\\ 
 \hline
 1400 & GaP        & 1.2$\times10^{-10}$ & 2.4$\times10^{-5}$ & 6.0$\times10^{6}$ & 7.3$\times10^{6}$ & 1.0$\times10^{6}$   & 2.0$\times10^{13}$ & 5.5$\times10^{6}$ & 1.1$\times10^{6}$ & 3.32\\ 
 \hline
 2800 & GaP        & 1.5$\times10^{-10}$ & 7.2$\times10^{-5}$ & 2.0$\times10^{6}$ & 9.8$\times10^{5}$ & 3.4$\times10^{5}$   & 2.4$\times10^{12}$ & 5.6$\times10^{5}$ & 2.0$\times10^{5}$ & 3.34\\ 
 \hline
\end{tabular}
% \end{adjustwidth}
\end{center}
\caption{Parameters used to calculate the cascaded output power for the ring cavity design in~\figref{ring eta abs} using a starting input power of $P_{\rm{in}}= 1000\,$mW.}
\label{tb: ring parameters}
\end{table}

\FloatBarrier

%%%%%%%%%%%%%%%%%%%%%%%%%%%%%%%%%%%%%%%%%%%%%%%%%%%%%%%%%%%%%%%%%%%%%%%%%%%%%%%%%%%%%%%%%%%%%%%%%
% NONLINEAR MATERIAL ANALYSIS
%%%%%%%%%%%%%%%%%%%%%%%%%%%%%%%%%%%%%%%%%%%%%%%%%%%%%%%%%%%%%%%%%%%%%%%%%%%%%%%%%%%%%%%%%%%%%%%%%

\section{Nonlinear material analysis }\applab{nonlinear material analysis}

We consider nonlinear optical susceptibilities in the frequency region near phonon polariton resonances (termed the Reststrahl band).
The coupling of a photon to a transverse optical (TO) phonon at frequency $\omega_{TO}$ in the THz or mid-IR yields the polariton dispersion relation and corresponding (linear) permittivity. Due to the shortage of high-power CW laser sources in the THz gap, however, there are few reports found in the literature of experiments measuring nonlinear optical susceptibilities near the Reststrahl region. As noted in Ref.~\cite{paarmann2015second}, which reported SHG experiments near the Reststrahl region of SiC, the few existing experimental studies~\cite{paarmann2015second,mayer1986far,dekorsy2003infrared,barmentlo1994sum,liu2008sum} are limited. 
Therefore, we rely on a theoretical model to estimate the nonlinear susceptibilities, but emphasize that further experiments are needed.

A harmonic oscillator-based model of the $\chi^{(2)}$ nonlinear susceptibility in the Reststrahl region of crystalline materials was developed by Garrett in 1968~\cite{garrett1968nonlinear}. Rather than electronic polarization, the dominant contribution in the region of polaritons comes from the transverse displacements of atoms in the lattice, which introduces a second, ionic, coordinate into the Lorentzian oscillator model~\cite{garrett1968nonlinear}. 
Garrett considered the "anharmonic vibronic oscillator", an extension of the electronic oscillator model used to calculate higher order susceptibilities by perturbation theory,
% and proved its equivalence to the quantum mechanical description.
and derived a generalized Miller's Rule for nonlinear susceptibilities when both electronic polarization and lattice deformation are taken into account.
% Consistent with this model are 
Faust and Henry carried out early THz difference frequency generation experiments in GaP and fit the data to an anharmonic oscillator model of the $\chi^{(2)}$ for III-V semiconductors~\cite{faust1966mixing}. Mayer and Keilmann used these models to analyze THz SHG experiments in GaAs and LiTaO$_3$~\cite{mayer1986far}. Here we use Garrett's generalized Miller's Rule for the ferroelectric materials (LiTaO$_3$ and LiNbO$_3$) and the Faust-Henry model for the zincblendes (GaAs, GaP, ZnTe). We briefly describe them below.
% \color{black}
%

The Lorentzian oscillator model describes the relative permittivity $\epsilon = (n-ic\alpha/2\omega)^2$ near lattice modes,
\begin{align}
    \epsilon = \epsilon_{\infty} + \sum_j \frac{S_j \omega_{j\text{TO}}^2}{\omega_{j\text{TO}}^2 - \omega^2 + i\gamma_{j\text{TO}} \omega},
    \label{eq:permittivity}
\end{align}
where $\omega_{jTO}$ is the resonant frequency of the $j$th transverse optical phonon, $\gamma_{jTO}$ is the linewidth of the resonance, $S_j$ is the strength of the oscillator, and $\epsilon_\infty$ is the high frequency limit of the  permittivity. Parameters used for each material are listed in Table~\ref{tb:parameters}. % Fig. \ref{fig:thz-params} plots the refractive index $n$ and absorption coefficient $\alpha$ obtained from this expression.

Perturbation theory relates the second order nonlinear susceptibility to the first order susceptibilities according to the anharmonic oscillator model \cite{boyd2003nonlinear}.
% \begin{align}
%     \chi^{(2)}(\omega_3,\omega_2,\omega_1) = \frac{\epsilon_0^2 m a}{N^2 e^3} \chi^{(1)}(\omega_3)\chi^{(1)}(\omega_2)\chi^{(1)}(\omega_1),
% \end{align}
% where $\epsilon_0$ is the vacuum permittivity, $m$ is the electron mass, $a$ is the potential anharmonicity, $N$ is the number density, and $e$ is the electron charge. 
Miller observed empirically that the ratio between the second order susceptibility and the products of the first order susceptibilities, often denoted as Miller’s $\delta$ coefficient, remains relatively constant across materials and frequencies \cite{miller1964optical}.
% 
% \color{blue}
% While it is generally understood that Miller's Rule is an approximation and questionable near resonances, experiments verify that SHG power is indeed enhanced near the TO phonon resonances of the material~\cite{paarmann2015second}
% \textcolor{red}{[and is supported by XXX theory]}.
% 
% \color{black}
Garrett extended the description into the frequency range near vibrational modes, to include both electronic and ionic contributions to the $\chi^{(2)}$ via a sum of terms \cite{garrett1968nonlinear}:
% \color{blue}
\begin{align}
    \chi^{(2)}_{jkl}(\omega_1,\omega_2, \omega_3) = \sum_{abc} \delta_{jkl}^{abc} \chi_j^a(\omega_1) \chi_k^b(\omega_2) \chi_l^c(\omega_3),
    \label{eq:miller-rule}
\end{align}
% \begin{align}
%     \chi^{(2)}_{jkl}(\omega,\omega, 2\omega) = \sum_{abc} \delta_{jkl}^{abc} \chi_j^a(\omega) \chi_k^b(\omega) \chi_l^c(2\omega) 
%     \label{eq:miller-rule}
% \end{align}
where the $\delta_{jkl}^{abc}$'s are generalized Miller's $\delta$'s, the labels $a,b,c$ are either $i$ (referring to the ionic coordinate) or $e$ (referring to the electronic coordinate), and the tensor component of the second order susceptibility is indexed by $jkl$. The $\chi_{j,k,l}^{a,b,c}$ are the \emph{linear} susceptibilities at the associated frequencies. %calculated using~\eqref{permittivity} with $\chi = \epsilon - 1$. In the case of SHG, $\omega_1=\omega_2\equiv\omega$ and $\omega_3 = 2\omega$.
% Near phonon resonances, the dielectric permittivity takes the form \cite{barker1970infrared}

When all three frequencies involved in the interaction are in the vicinity of TO phonon resonances, that is, $\omega_1 \approx \omega_2 \approx \omega_3$, the denominator of~\eqref{permittivity} is resonant and we can make the approximation $\chi^i \gg \chi^e$. Further assuming that the $\delta_{jkl}^{abc}$'s are comparable in magnitude (which is typically within reason), we can retain just the last term \cite{garrett1968nonlinear}
\begin{align}\label{eq:chi2-3m supp}
    \chi^{(2)}_{jkl}(\omega_3,\omega_2,\omega_1) = \delta_{jkl}^{iii} \chi_j^i(\omega_3) \chi_k^i(\omega_2) \chi_l^i(\omega_1),
\end{align}
% \begin{align}\label{eq:chi2-3m supp}
%     \chi^{(2)}_{jkl}(\omega,\omega,2\omega) \approx \delta_{jkl}^{iii} \chi_j^i(\omega) \chi_k^i(\omega) \chi_l^i(2\omega) 
% \end{align}
% 
% where $\delta_{jkl}^{iii}$ is the ionic Miller's coefficient and the $\chi^i(\omega)$ are the linear susceptibilities evaluated at the frequencies $\omega, 2\omega$ in the THz range. 
where $\chi^i$ is calculated using~\eqref{permittivity} with $\chi^i = \epsilon - 1$ (ordinary or extraordinary permittivity is chosen depending on the relevant tensor component).
Comparison with the expression above for permittivity shows that the $\chi^{(2)}$ is a product of oscillator terms which retains its resonant features. For LiTaO$_3$ and LiNbO$_3$, which have high permittivities even at frequencies well below the TO phonon resonances ($\epsilon = 37.6$ and $26.0$ along the c-axis, respectively~\cite{barker1970infrared,barker1967dielectric}) the approximation can still be made. (In general, however, all the terms in the sum must be considered). As reported values are lacking for $\delta^{iii}_{311}$ and $\delta^{iii}_{333}$ of LiNbO$_3$, and $\delta^{iii}_{333}$ of LiTaO$_3$, we assume that as the frequencies approach the microwave limit, the susceptibilities should approach the experimental microwave values, and we use these to fit the $\delta$'s~\cite{boyd1971microwave}. With this fit, we obtain $\delta_{31}^{iii} = 25\times10^{-14}$ m/V for LiTaO$_3$, which we compare to $\delta_{31}^{iii} = 8\times10^{-14}$ m/V reported in Ref.~\cite{mayer1986far}.
Additionally, we find $\delta_{33}^{iii} = 50\times10^{-14}$ m/V for LiTaO$_3$, $\delta_{31}^{iii} = 36\times10^{-14}$ m/V and $\delta_{33}^{iii} = 85\times10^{-14}$ m/V for LiNbO$_3$.

% \textcolor{red}{[LA - add to this]} We use this model to calculate the second order susceptibilities in these ferroelectric materials. We estimate the $\delta^{iii}_{311}$ and $\delta^{iii}_{333}$ by fitting the low-frequency limit of $\chi^{(2)}$ to experimentally determined microwave nonlinear coefficients~\cite{S_boyd1971microwave,S_boyd1973microwave}.

% In $\bar{4}3m$ crystals, 
% the approximation to Miller's rule does not apply. Instead, 
% the dispersion of $\chi^{(2)}$ for SHG is described by the Faust-Henry model (see references for more general expressions for three-wave mixing) \cite{S_faust1966mixing}:
In the zincblende crystals (GaAs, GaP, ZnTe),
the dispersion of $\chi^{(2)}$ is instead described by the Faust-Henry model \cite{faust1966mixing}:
% 
% \begin{align}\label{eq:chi2-43m supp}
%     \chi^{(2)}_{jkl}(\omega_3,\omega_2, \omega_1) &= \chi_E^{(2)} \left[ 1+C_1 \left( \frac{1}{D(\omega_1)}+\frac{1}{D(\omega_2)}+\frac{1}{D(\omega_3)} \right) \right] 
% \end{align}
% 

% Several authors have proposed phenomenological models of the nonlinear susceptibilities at low frequencies.  For crystals with zinc-blende structure, the expression for $\chi^{(2)}$ is given by the Faust-Henry model \cite{faust1966mixing, flytzanis1969second}

\begin{align}
\begin{split}
    \chi^{(2)}(\omega_3,\omega_2, \omega_1) &= \chi_E^{(2)} \Bigg[ 1+C_1 \left( \frac{1}{D(\omega_1)}+\frac{1}{D(\omega_2)}+\frac{1}{D(\omega_3)} \right)  \\
    &+ C_2\left(\frac{1}{D(\omega_1)D(\omega_2)}+\frac{1}{D(\omega_1)D(\omega_3)}+\frac{1}{D(\omega_2)D(\omega_3)} \right) + C_3\frac{1}{D(\omega_1)D(\omega_2)D(\omega_3)} \Bigg] 
\end{split}
\end{align}

% \begin{align}
% % \begin{split}
%     \chi^{(2)}(\omega,\omega, 2\omega) &= \chi_E^{(2)} \Bigg[ 1+C_1 \left( \frac{2}{D(\omega)}+\frac{1}{D(2\omega)} \right) 
%     + C_2\left(\frac{1}{D^2(\omega)}+\frac{2}{D(\omega)D(2\omega)}\right) + C_3\frac{1}{D^2(\omega) D(2\omega)} \Bigg] 
% % \end{split}
% \end{align}
% Faust-henry equation
where $\chi^{(2)}_E$ is the electronic nonlinear susceptibility, $C_{1,2,3}$ are the Faust-Henry coefficients, and 
$D(\omega) = 1-(\omega/\omega_\text{TO})^2 - i\gamma_\text{TO}\omega/\omega_\text{TO}^2$. 
% $D(\omega_n) = 1-(\omega_n/\omega_{TO})^2 - i\gamma_{TO}\omega_n/\omega_{TO}^2$. 
% A critical advantage comes from the resonances in Eqs. \ref{eq:chi2-3m} and \ref{eq:chi2-43m}.
The $C_1$ term accounts for interactions with one frequency near the TO phonon resonance, the $C_2$ term for interactions with two frequencies nearby, and the $C_3$ term accounts for all three contributions near the phonon resonance. Values (where available in the literature) can be found in Table \ref{tb:zincbl-params}.

% \color{blue}
We note that while the nonlinear optical properties of solids may be calculated more accurately from first principles via alternative methods such as density functional theory and bond charge models~\cite{levine1973bond,levine1991calculation,wang1999calculation,zhang2010first,ratnaparkhe2020calculated}, such calculations are typically complex (relying on many structural parameters) and therefore computationally expensive. 
While not a substitute for a full quantum mechanical treatment, the classical phenomenological models are simple and have proven useful in fitting experimental data and estimating order of magnitude parameters.
Since the overall conversion efficiency of the device depends on both material and geometric factors, discrepancies in the estimate of $\chi^{(2)}$ can be offset by modified cavity designs. Furthermore, phase matching only requires accurate estimates of the refractive index, which have been well-studied by experiments. In practice, thermal or electro-optic tuning may be used to correct for fabrication imperfections. Thus, for our purposes, the phenomenological treatment we present is sufficient. Nonetheless, the potential applications and impacts motivate further experimental studies of nonlinear optical material properties in the THz/mid-IR spectrum to verify the predictions posed here. 
\color{black}

\begin{table}[htbp!]
\begin{center}
\begin{tabular}{| c | c | c | c | c | c | c |}
\hline
 \textbf{Material} & \textbf{Pol.} & \textbf{$\epsilon_\infty$} & \textbf{$\Delta\epsilon$} & \textbf{$\omega_{TO}$ (THz)} & \textbf{$\gamma_{TO}$ (THz)} & \textbf{Ref(s).} \\ 
 \hline
 GaAs & - & 11.55 & 1.95 & 8.05 & 0.29 & Refs.~\cite{dekorsy2003infrared, mayer1986far} \\ 
 \hline
 GaP & - & 9.09 & 1.92 & 10.94 & 0.11 & \cite{faust1966mixing, barker1968dielectric} \\
\hline
ZnTe & - & 3.0 & 6.92 & 5.3 & 0.09 & \cite{carnio2020extensive, hattori1973indices} \\
\hline
\multirow{10}{4em}{LiTaO$_3$} & \multirow{6}{3em}{ord} & \multirow{6}{3em}{7.73} & 24.10 & 4.26 & 0.42 & \multirow{6}{3em}{\cite{barker1970infrared}} \\
& & & 2.40 & 7.59 & 0.27 & \\
& & & 2.50 & 9.47 & 0.42 & \\
& & & 2.00 & 11.24 & 0.78 & \\
& & & 0.03 & 13.85 & 0.18 & \\
& & & 2.33 & 17.81 & 0.96 & \\
\cline{2-7}
& \multirow{4}{3em}{ext} & \multirow{4}{3em}{6.78} & 30.00 & 6.00 & 0.84 & \multirow{4}{3em}{\cite{barker1970infrared}}\\
& & & 0.005 & 10.70 & 0.33 & \\
& & & 2.66 & 17.87 & 0.54 & \\
& & & 0.34 & 19.70 & 1.68 & \\
\hline
\multirow{13}{4em}{LiNbO$_3$} & \multirow{8}{3em}{ord} & \multirow{8}{3em}{5.02} & 22.0 & 4.56 & 0.42 & \multirow{8}{3em}{\cite{barker1967dielectric}}\\
& & & 0.80 & 7.08 & 0.36 & \\
& & & 5.50 & 7.94 & 0.36 & \\
& & & 2.20 & 9.65 & 0.33 & \\
& & & 2.30 & 10.88 & 0.99 & \\
& & & 0.18 & 12.92 & 0.36 & \\
& & & 3.30 & 17.57 & 1.05 & \\
& & & 0.20 & 20.09 & 1.41 & \\
\cline{2-7}
& \multirow{5}{3em}{ext} & \multirow{4}{3em}{6.16} & 16.00 & 7.44 & 0.63 & \multirow{4}{3em}{\cite{barker1967dielectric}} \\
& & & 1.00 & 8.21 & 0.42 & \\
& & & 0.16 & 9.20 & 0.75 & \\
& & & 2.55 & 18.83 & 1.02 & \\
\hline
\end{tabular}
\end{center}
\caption{Parameters used to calculate THz refractive indices and absorption coefficients in Fig. 2 of the main text.}
\label{tb:parameters}
\end{table}

\begin{table}[htbp!]
\begin{center}
\begin{tabular}{| c | c | c | c | c | c |}
\hline
\textbf{Material} & $\chi^{(2)}_E$ (pm/V) & $C_1$ & $C_2$ & $C_3$ & \textbf{Ref.} \\ \hline 
GaAs & 268 & -0.59 & 0.14 & -0.07 & \cite{dekorsy2003infrared} \\ \hline 
GaP & 156 & -0.53 & - & - & \cite{faust1966mixing} \\ \hline 
ZnTe & 139 & -0.07 & - & - & \cite{carnio2020extensive} \\ \hline 

\end{tabular}
\end{center}
\caption{Parameters used to calculate second-order nonlinear susceptibility of the zincblende materials, where available.}
\label{tb:zincbl-params}
\end{table}

%%%%%%%%%%%%%%%%%%%%%%%%%%%%%%%%%%%%%%%%%%%%%%%%%%%%%%%%%%%%%%%%%%%%%%%%%%%%%%%%%%%%%%%%%%%%%%%%%
% CAVITY DESIGNS
%%%%%%%%%%%%%%%%%%%%%%%%%%%%%%%%%%%%%%%%%%%%%%%%%%%%%%%%%%%%%%%%%%%%%%%%%%%%%%%%%%%%%%%%%%%%%%%%%

\section{Cavity designs}\applab{app cavity designs}

\subsection{PhC cavities}\applab{PhC cavity designs}
\begin{figure}[htbp!]
    \centering
    \includegraphics[width=0.6\textwidth]{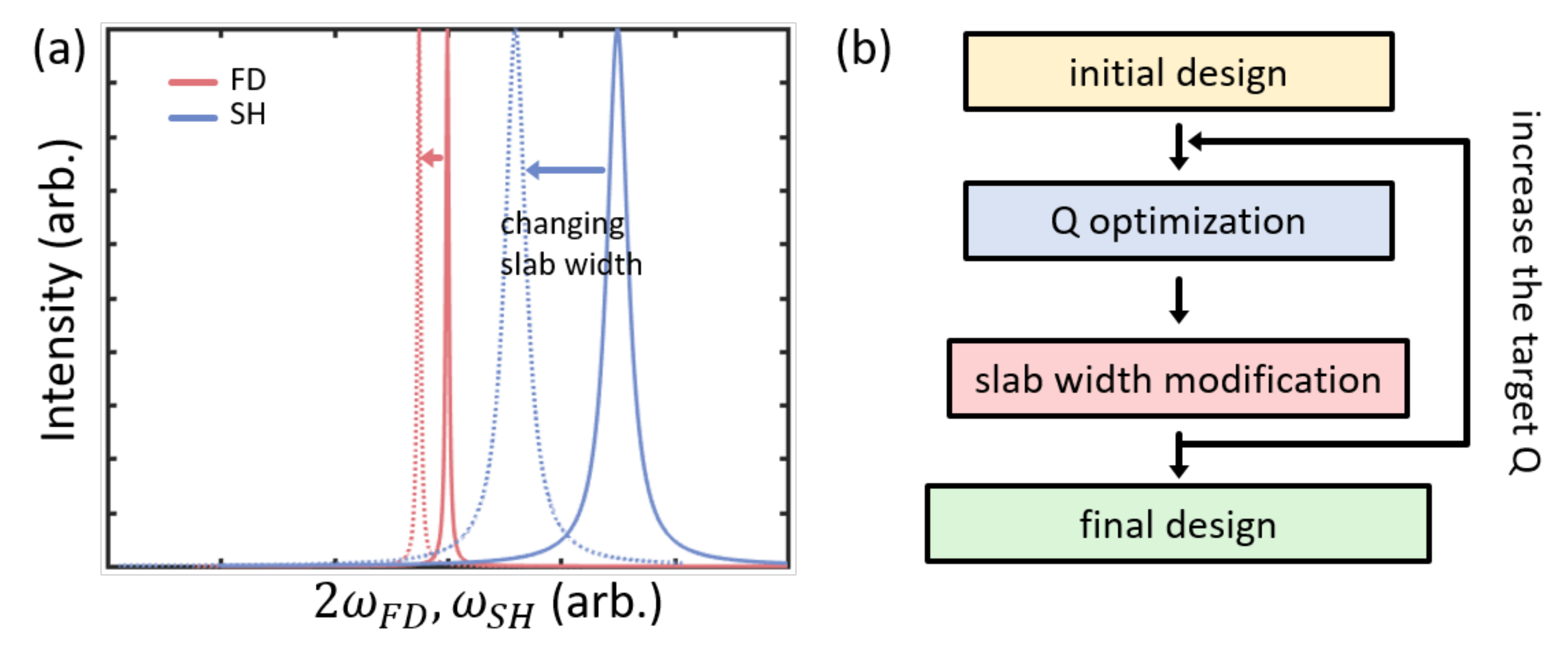}
    \caption{A design flow for doubly resonant photonic crystal cavity. (a) The effect of changing the slab width. Solid lines represent the original cavity and dashed lines represent the resulting spectra after changing the slab width. (b) Flow chart describing our design algorithm. We first make an initial design that has two resonances with $\omega_b\approx 2\omega_a$, but with $\omega_b- 2\omega_a$ being larger than the resonance linewidth. We then alternate between $Q$-optimization and slab width modification. At each iteration, we increase the target $Q$ while gradually approaching the resonance condition $\omega_b = 2\omega_a$.}
    \label{fig:design_flow}
\end{figure}
Photonic crystal cavities have two advantages in SHG: 1) they have a small mode volume resulting in increased energy density (SHG is proportional to the intensity squared); 2) they are phase-matching insensitive because the nonlinear material is only at the center. However, it is often very hard to obtain the energy matching condition ($\omega_b = 2\omega_a$) with PhC cavities. In this section, we show a systematic design approach for doubly resonant PhC cavities. 

For a dielectric defect cavity, where the defect pulls down the air-like mode into the bandgap, often there are higher-order localized modes. We assume the following general situation; we have an initial cavity design with two low-$Q$ cavity modes; they are near second-harmonic, $\omega_b = 2\omega_a + \delta$, but with large detuning, $\delta>\kappa_a,\kappa_b$. In this case, the conversion efficiency is small due to small $Q$-factors and large detuning. The goal of our design is to 1) increase $Q_a$ and 2) $Q_b$ and to 3) reduce the detuning, $\delta\ll\kappa_a, \kappa_b$. A direct optimization of three conditions requires the exploration of a full parameter space, which is time-intensive.

Alternatively, we decoupled the optimization of the three conditions by separating the parameter space into three. The separation was chosen to be only sensitive to the corresponding condition to be optimized. More specifically, we optimized $Q_b$ by changing the radii and positions of the first seven holes on each side of the center hole with the tips. On the other hand, optimization of $Q_a$ happens through changing the outer twenty-five holes. Out of the twenty-five, the latter five have the radii and distances fixed to form Bragg mirrors. Our rationale for the choice of separation is that the fundamental mode is less tightly confined than the second-harmonic one. Thus, the optimization of the fundamental mode minimally affects the second-harmonic mode. 

On the other hand, we adjust the detuning, $\delta=\omega_b-2\omega_a$, with the slab width (Fig.~\ref{fig:design_flow}). The second-harmonic mode we chose is the third order waveguide mode in the slab-width direction (for the mode anti-node to be located at the tip). Comparing with the fundamental mode (the first order waveguide mode), the second-harmonic mode is more tightly confined in the transversal direction (larger transverse $k$-vector). Thus, the resonant frequency of the second-harmonic mode is more sensitive to the slab width. 

Our optimization process is summarized in Fig.~\ref{fig:design_flow}(b). We perform the three optimizations and iterate. Starting with the initial design, $Q_a$ and $Q_b$ are optimized sequentially. When the target $Q$-factors are achieved, the width of the slab is adjusted to reduce the detuning. Despite separating the parameter space to minimize the coupling between parameters, the modification of the slab width reduces the $Q$s. We then increase the target $Q$s and iterate the optimization. When the $Q$s exceed the target value, we finalize the design.

% \color{blue}
\subsection{PhC cavity simulations}\applab{PhC cavity simulations}
% \color{black}
We simulated PhC cavities with finite-difference time-domain (FDTD) method using a commercial software, Lumerical. We simulated a cavity at 331 GHz (fundmental) and 662 GHz (second-harmonic) and used $Q$ and scaled mode field profiles at other frequencies (Maxwell's equations have scale invariance). The mode field profiles are obtained with LiNbO$_3$ as a nonlinear material ($n = 5.102$) at the center.

We used ($15~\mu$m$\times 15~\mu$m$\times 20~\mu$m) mesh and overrode the tip area ($100~\mu$m$\times 100~\mu$m$\times 139~\mu$m) with a mesh having a resolution of $500~$nm$\times 500~$nm$\times 10~\mu$m. The cavity is first excited with a dipole, and the target modes are extracted using a standard mode profile monitor. We used the extracted mode for subsequent simulations for fast convergence and accurate parameter extraction. The quality factors of cavities are deduced from the decay rate of the electric field envelope (high-Q analysis).

\subsection{Ring cavities}\applab{Ring cavity designs}

\begin{figure}[htbp!]
    \centering
    \includegraphics[width=\textwidth]{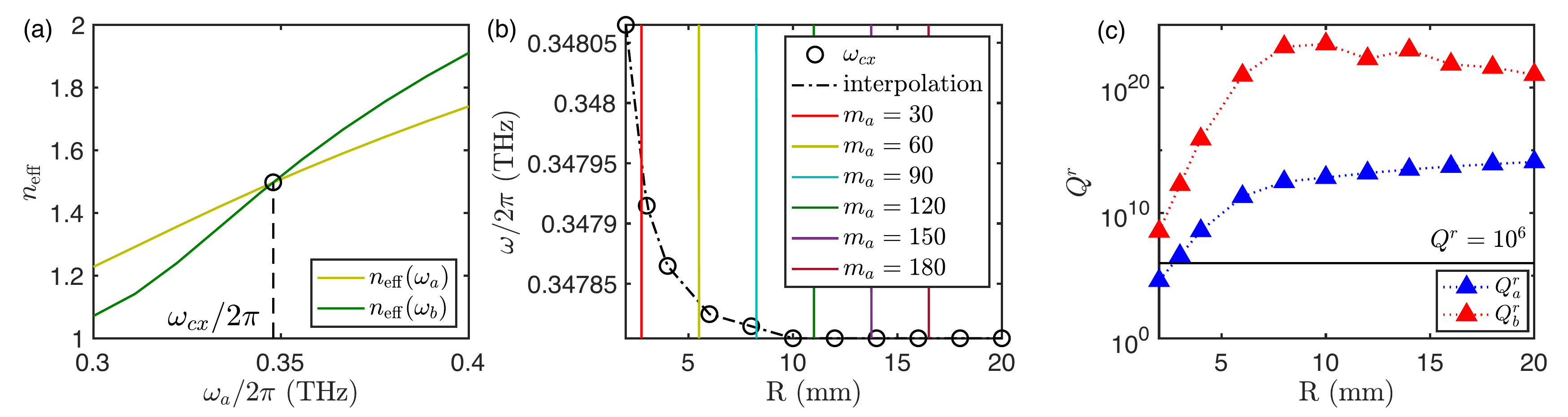}
    \caption{Phase matching and radiation quality factor of the ring. (a) Effective index curves for fundamental and second harmonic modes of LiNbO$_3$ ring cavity with 100 nm $\times$ 100 nm tip area, shown for an example radius of $3$ mm. (b) Waveguide $n_\text{eff}$ crossover frequency plotted against radius and ring resonant frequencies with associated radius and $n_\text{eff}$ for several integer values of $m_a$. Intersections of the colored curves with $\omega_{cx}$ give parameters at which the ring is phase matched ($n_\text{eff}(\omega_a) = n_\text{eff}(2\omega_a)$) and resonant at both $\omega_a$ and $2\omega_a$. (c) $Q_{a(b)}^r$ due to bending loss calculated for different $R$.}
    \label{fig:ring_pm}
\end{figure}

\textbf{Phase-matching.} Maximizing the conversion efficiency requires the ring to be both phase-matched and doubly resonant at the fundamental and SH frequencies. For two resonant frequencies of the ring, $\omega_{a(b)}$, the resonance condition is 
$\omega_{a(b)} = \frac{m_{a(b)} c}{R n_\text{eff}(\omega_{a(b)},R)}$,
% \begin{align}
%     % \frac{R \omega_{a(b)} n_\text{eff}(\omega_{a(b)},R)} {c} = m_{a(b)},
%     \omega_{a(b)} = \frac{m_{a(b)} c}{R n_\text{eff}(\omega_{a(b)},R)},
% \end{align}
where $m_{a(b)}$ is 
% the integer wavenumber of the ring
an integer that enumerates the azimuthal modes, $n_\text{eff}(\omega_{a(b)},R)$ is the effective index of the waveguide, and $R$ is the radius. With $\omega_b = 2\omega_a$,  the phase-matching requirement $m_b = 2m_a$  is satisfied when $n_\text{eff}(\omega_a,R) = n_\text{eff}(2\omega_a,R)$. Using a finite difference eigenmode solver, we compute the effective indices of the fundamental and second harmonic modes as a function of $\omega_a$ and $R$. The waveguide parameters are adjusted such that the effective index of the fundamental and SH modes are equal at a crossover frequency $\omega_a=\omega_\text{cx}$, shown in~\figref{ring_pm}(a) for LiNbO$_3$ near 350 GHz. 
% This first requires a waveguide cross section design that has a crossover in the $n_\text{eff}$ curves, which is found by sweeping the cross-section parameters (i.e. width, height).
As the bending radius is reduced, the crossover frequency $\omega_\text{cx}$ increases, as plotted in~\figref{ring_pm}(b). Where $\omega_\text{cx}$ intersects with one of the ring resonances (colored lines in~\figref{ring_pm}(b)), the ring is  simultaneously doubly resonant and phase-matched.
Assuming the crossover frequency and effective index are well-behaved functions, they can be interpolated to find appropriate radii for ring resonance.
For simplicity, in the main text we use a straight waveguide to approximate the efficiencies; however, with this method we confirm that the mode profiles and parameters are comparable to the bent waveguide case, and in principle can be used to phase match for a desired material and waveguide design. 

% To design for operation at different frequencies, the host waveguide dimensions are scaled uniformly, while the tip size is held constant at a cross-sectional area of $s = 100$ nm $\times 100$ nm.

\textbf{Radiation quality factor.} Radiation loss limits the ring radius. To account for radiation losses due to bending of the ring, we calculated the radiation quality factor $Q_{a(b)}^r$  from the imaginary part of the modal effective index. \figref{ring_pm}(c) shows the minimum radius that is compatible with a radiation quality factor, $Q_{a(b)}^r \geq 10^6$. Considering this and the phase-matching condition, we choose a radius of $R = 3.2$ mm.

\textbf{Material quality factor.} To calculate the material quality factor $Q_{a(b)}^m$, we treated the absorption due to a small region of lossy nonlinear material as an imaginary perturbation to the dielectric function (see~\appref{material absorption} for details). 
The material quality factor is given by $Q_{a(b)}^m = \frac{\omega}{2\tilde{\kappa}_{a(b)}^m} = \frac{\epsilon_{a(b)}'}{\sigma_{a(b)} \epsilon_{a(b)}''}$
where the real and imaginary parts of the relative permittivity $\epsilon_{a(b)} = \epsilon_{a(b)}' + i\epsilon''_{a(b)}$ are given by the Lorentz oscillator model described in the main text, and $\sigma_{a(b)}$ is an overlap integral defined in~\eqref{material loss}.

In Fig. 4 of the main text, we assumed critical coupling, which maximizes the SHG efficiency in the nondepleted pump limit. The overall quality factor $1/Q_{a(b)} = 1/Q_{a(b)}^c + 1/Q_{a(b)}^m + 1/Q_{a(b)}^r$ was used to compute the SHG efficiency using~\eqref{eta SHG}.

\textbf{Nonlinear mode overlap.} 
The SHG overlap $\tilde{\beta}$ can be approximated in the ring resonator by \cite{guo2016second}
\begin{align}%\label{eq:beta}
    \tilde{\beta}&^{ring} \approx \frac{1}{\sqrt{2\pi R}}
    \frac{\int \bar{\epsilon}(x,y) E^2_{a,y}(x,y)E^*_{b,y}(x,y) dx dy}{\int \epsilon_a(x,y)\left|\vec{E}_a(x,y)\right|^2 dx dy \sqrt{\int \epsilon_b(x,y)\left|\vec{E}_b(x,y)\right|^2 dx dy}}
    \sqrt{\frac{\lambda_a^3}{n_a^3}}\delta(m_b-2m_a),
\end{align}
%
%Where $\int_C$ is taken over the entire cavity, $\int_M$ is taken only over the nonlinear material, 
where $\bar{\epsilon}(x,y)$ is a function that equals 1 inside the nonlinear material and 0 everywhere else,
and we assume coupling of three TM modes ($E$ field polarized in the $y$ direction) by the $\chi^{(2)}_{33}$ component of the nonlinear material $(\chi^{(2)}_\text{eff} = \chi^{(2)}_{33})$. The SHG overlap is maximized when the phase matching condition  $\delta(m_b-2m_a)$ is satisfied (with $\delta$ denoting the Dirac delta function).

\FloatBarrier

\subsection{Ring cavity simulations} \applab{Ring cavity simulations}

% 
% \color{blue}
Mode field profiles of the ring resonator were computed using the commercial 2D Finite Difference Eigenmode (FDE) solver Lumerical MODE and the open-source software MPB. 
% \color{black}
% Mode profiles of the ring resonator were obtained using a commercial 2D Finite Difference Eigenmode (FDE) solver (Lumerical MODE). 
At $\omega_a/2\pi \approx$ 350 GHz, the dimensions of the ring resonator cross section as indicated in Fig. 3 of the main text are
$w$ = \SI{150}{\micro\metre}, $h_a$ = \SI{30}{\micro\metre}, $h_b$ = \SI{91.8}{\micro\metre}, $w_{br}$ = \SI{20}{\micro\metre}, $w_{tip}$ = 625 nm, and $h_{tip}$ = 625 nm. For calculation of the efficiencies in Fig. 4, however, we used $w_{tip} = h_{tip} = 100$ nm across all frequencies, which is in principle achievable with standard photolithography while being large enough relative to the lattice constants to treat the material as a continuous medium. 
% w=150um,ha=30um,hb=91.8um,w_br=20 μm,w_tip=100 nm,  and htip=100 nm. 
%The ring radius used to calculate radiation loss is 3.2 mm. The frequency was swept to find an intersection between the effective mode indices of the FD and SH modes. (See discussion on phase matching in Supplementary Section 3C.) The simulation was then repeated at the crossover frequency to obtain the mode profiles. For the other frequencies in the cascade, all dimensions except $w_{tip}$  and $h_{tip}$ were scaled proportionately to frequency (neglecting the dispersion of the nonlinear material for simplicity). 

% \color{blue}
Field profiles in Fig. 3b were obtained using MPB for a straight waveguide with a uniform mesh of 122 nm $\times$ 122 nm at 350 GHz across the entire simulation region and periodic boundary conditions. To resolve the smallest tip sizes without prohibitive memory requirements, however, nonuniform meshing was used (available in Lumerical MODE but not MPB). Therefore, all calculations in Figs. 3c-f and Fig. 4 were implemented in Lumerical by applying a nonuniform mesh across the simulation region to ensure sufficient resolution in the nonlinear tip (minimum mesh size of 5 nm $\times$ 5 nm at 350 GHz). 
% \color{black}

% \color{blue}
To calculate radiation losses (in Lumerical), a bending radius was applied and perfectly matched layer (PML) boundary conditions were used on the three outer sides of the ring cross section; the fourth side corresponding to the direction of the center used metal boundary conditions. The bounding box was chosen to be large enough for convergence of the radiation $Q$ to eliminate artifacts due to the PML.

The integrals in the expressions for $\tilde{\beta}^{ring}$ and $\sigma_{a(b)}$ were numerically evaluated over the entire (2D) simulation region. 

\subsection{Material absorption} \applab{material absorption}
Let us assume that we have data for the material loss from a measurement of the decay of an electromagnetic field as it propagates through the material. We assume the experimental conditions are such that we may consider the field to be a plane wave that evolves as $\exp(ik_0 \tilde{n} z - i\omega_0 t)$. Then, the complex refractive index is defined as
\begin{align}
     \tilde{n} \equiv n + in'' &\equiv \sqrt{\epsilon' + i \epsilon''} = \sqrt{\epsilon'}\sqrt{1 + \frac{\epsilon''}{\epsilon'}} \approx \sqrt{\epsilon'} \left( 1 + i\frac{\epsilon''}{2\epsilon'} \right) \Rightarrow \nn \\
     n &= \sqrt{\epsilon'}, ~~\text{and} ~~ n'' = \frac{\epsilon''}{2n},
\end{align}
where we assumed $\epsilon'\gg \epsilon''$. The intensity absorption coefficient is
\begin{align}
     \alpha \equiv 2 k_0 n'' = \frac{\omega_0}{c} \frac{\epsilon''}{n},
\end{align}
where $k_0=\omega_0/ c$ is the free-space wave number. If the intensity of the wave propagates at the group velocity and decays with the amplitude decay rate $\kappa$, then we have
\begin{align}\eqlab{alpha vs kappa}
     \kappa \equiv \frac12 \alpha v_g = \frac12 \omega_0 \frac{\epsilon''}{n_g n} \approx \frac12 \omega_0 \frac{\epsilon''}{\epsilon'}.
\end{align}
We set $n_g\approx n$ in~\eqref{alpha vs kappa} since we assume the experiment was performed with an approximately plane wave propagating in a uniform material.\\

The contribution to material absorption from the nonlinear material inside our hybrid cavity designs are found using perturbation theory~\cite{choi2017self, joannopoulos2008molding} 
% 
% \begin{align}\eqlab{material loss}
%      \widetilde{\epsilon_{a(b)}''} = \epsilon_{a(b)}'' \frac{\int dV \bar{\epsilon}(\bm{r}) |\vec{E}_{a(b)}(\vec{r})|^2  }{ \int dV \epsilon(\bm{r}) |\vec{E}_{a(b)}(\vec{r})|^2} \equiv \epsilon_{a(b)}'' \sigma_{a(b)} ~\Rightarrow ~~ \widetilde{\kappa_{a(b)}^m} \equiv \kappa_{a(b)}^m \sigma_{a(b)},
% \end{align}
\begin{align}\eqlab{material loss}
     \tilde{\epsilon}_{a(b)}'' = \epsilon_{a(b)}'' \frac{\int dV \bar{\epsilon}(\bm{r}) |\vec{E}_{a(b)}(\vec{r})|^2  }{ \int dV \epsilon(\bm{r}) |\vec{E}_{a(b)}(\vec{r})|^2} \equiv \epsilon_{a(b)}'' \sigma_{a(b)} ~\Rightarrow ~~ \tilde{\kappa}_{a(b)}^m \equiv \kappa_{a(b)}^m \sigma_{a(b)},
\end{align}
where $\bar{\epsilon}(\bm{r})$ is a function that equals 1 inside the nonlinear material and 0 everywhere else. Here, we denoted the effective decay rate using a $\widetilde{}$ and~\eqref{material loss} shows that the bulk decay rate, $\kappa_{a(b)}^m$, is reduced by a factor of $\sigma_{a(b)}$. Note, however, that elsewhere in the manuscript we use $\kappa_{a(b)}^m$ to denote the effective decay rate for notational convenience.  

\subsection{Material loss of host cavities} 
\begin{figure}[!htbp]
    \centering
    \includegraphics[width=0.75\textwidth]{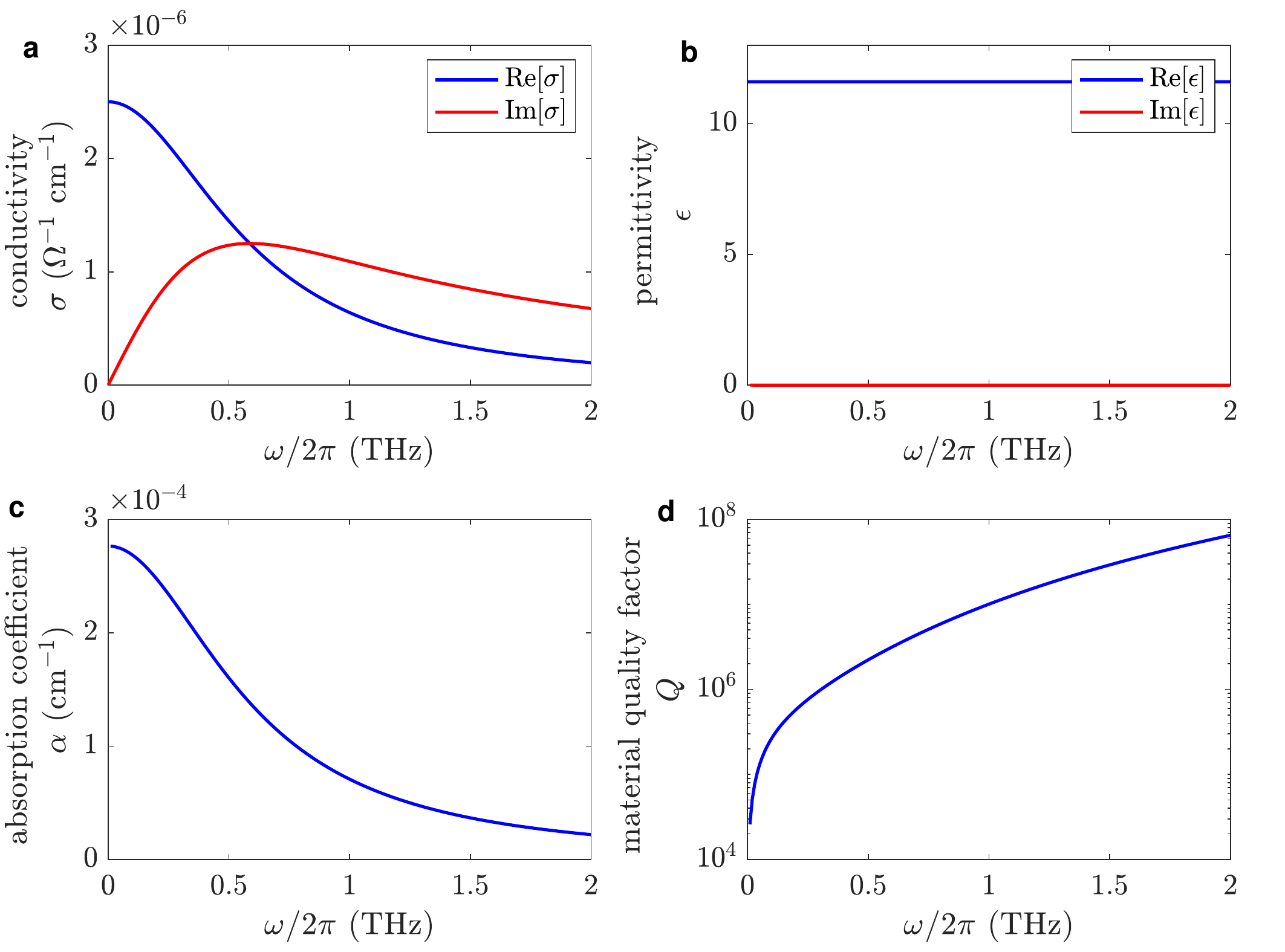}
    \caption{High-resistivity Si Thz characteristics with Drude model. (a) Conductivity, (b) permittivity, (c) absorption coefficient, and (d) material quality factor.}
    \label{fig:Si_Q}
\end{figure}

\applab{host material loss}
Here we discuss the material losses in the host cavities. While we assumed lossless silicon in the calculations, high-resistivity float-zone silicon (HRS) is a realistic candidate for the low-loss dielectric material~\cite{dai2004terahertz}. HRS is one of the conventionally used materials for THz and mm-wave applications because of its low absorption. Using the Drude model, 400 k$\Omega$ cm Si~\cite{krupka2016high,van1990carrier} has material quality factors $Q\sim4\times10^5$ at 150 GHz and $Q \geq 1\times10^6$ above $\sim$300 GHz (Fig.~\ref{fig:Si_Q}). 
We expect that an improvement by a factor of $\sim$20 in the resistivity of silicon would make it effectively lossless in all calculations, with $Q>10^7$.
% \color{red} We expect 20 times increase of the resistivity of silicon effectively makes silicon lossless in all calculations with $Q>10^7$ \color{black}

%For example, HRS with the resistivity of 10 k$\Omega$ cm has a peak absorption coefficient of $\alpha \sim$ 0.01 cm$^{-1}$, corresponding to material quality factors of $Q \sim 10^3 - 10^4$~\cite{S_dai2004terahertz}. To neglect these losses in SHG with the cavity designs we consider here, we require $Q \gtrsim 10^5 - 10^6$, but this can theoretically be achieved by increasing the resistivity of the HRS, since the absorption is inversely proportional to the resistivity as the introduction of carriers increases loss. Indeed, dielectric resonators made from 70 k$\Omega$ cm and 400 k$\Omega$ cm HRS have been shown to achieve $Q \sim 1.8\times10^4$ and $Q \sim 6\times10^4$, respectively, at 50 GHz~\cite{S_krupka2016high}. Using the Drude model for the dielectric permittivity with electronic losses, we predict the frequency dependent $Q$ for 400 k$\Omega$ cm HRS to exceed $2\times10^5$ above $\sim$100 GHz and $1\times10^6$ above $\sim$300 GHz (see Fig.~\ref{fig:Si_Q}; parameters are scaled from Ref.~\cite{S_van1990carrier}). Thus for simplicity, we neglect the frequency dependent losses due to material absorption in the host cavity. 
%

%%%%%%%%%%%%%%%%%%%%%%%%%%%%%%%%%%%%%%%%%%%%%%%%%%%%%%%%%%%%%%%%%%%%%%%%%%%%%%%%%%%%%%%%%%%%%%%%%
% DIELECTRIC BREAKDOWN
%%%%%%%%%%%%%%%%%%%%%%%%%%%%%%%%%%%%%%%%%%%%%%%%%%%%%%%%%%%%%%%%%%%%%%%%%%%%%%%%%%%%%%%%%%%%%%%%%

\section{Dielectric Breakdown} \applab{dielectric breakdown}

Due to the strong field confinement of our cavity designs, it is important to consider the limitations imposed by dielectric breakdown. To evaluate the field strength in the cavities, we re-normalize the eigenmodes as
\begin{align}\eqlab{mcE def }
     \vec{\mathcal{E}}_n(\vec{r}) \equiv \vec{E}_n(\vec{r}) \sqrt{\frac{\hbar\omega_n |\alpha_n|^2}{\int \epsilon_0 \epsilon(\vec{r}) |\vec{E}_n(\vec{r})|^2 dV }},
\end{align}
where $\hbar \omega_n|\alpha_n|^2$ is the energy in cavity mode $n$ found from~\eqref{eoms classical}. This re-normalized field correctly accounts for the electromagnetic energy in the cavity
\begin{align}\eqlab{mcE energy}
     \int \epsilon_0 \epsilon(\vec{r}) |\vec{\mathcal{E}}_n(\vec{r})|^2 dV = \hbar\omega_n |\alpha_n|^2,
\end{align}
and therefore provides the electric field in SI units of V/m.
Comparison of the maximum electric field in each cavity material with the dielectric strength of the corresponding material provides the limit on the input power given by dielectric breakdown.

Figures~\ref{fig:phc Emax} and ~\ref{fig:ring Emax} plot the maximum field strength in V/m for the PhC and ring, respectively, corresponding to the intra-cavity energies in Figs.~\ref{fig:phc eta abs} and~\ref{fig:ring eta abs} for $P_{in} = 200$ mW (blue), 400 mW (red), 1000 mW (green), and 4000 mW (black). 
%
% Figs.~\ref{fig:phc Emax}(b) and ~\ref{fig:ring Emax}(b) 
Parts (b) indicate that the maximum electric field in the air near the tips exceeds the breakdown field of $3\times10^6$ V/m~\cite{berger2006dielectric}; operating the device in vacuum would shift the consideration onto the waveguide materials. 
Figs.~\ref{fig:phc Emax}(a) and ~\ref{fig:ring Emax}(a) show that the maximum electric field in the silicon peaks near $1\times10^8$ V/m in both the PhC and the ring, and that in the nonlinear material peaks near $6\times10^7$ V/m in the PhC (lithium niobate) and $1.5\times10^8$ V/m in the ring (gallium phosphide), for 4 W of input power at the seed oscillator.
For comparison, gallium phospide exhibits a breakdown field of $1\times10^8$ V/m~\cite{shur1996handbook}, while
bulk lithium niobate can tolerate THz field strengths of over $1\times10^8$ V/m without breakdown~\cite{hirori2011single}, and possibly higher for the small tip gaps considered here. Silicon, on the other hand, has a theoretical intrinsic breakdown field of $8\times10^7$ V/m (independent of electrode separation), making it the limiting material in this case. Issues with dielectric breakdown can be circumvented by using materials with higher breakdown fields (e.g. diamond, with $2\times10^9$ V/m~\cite{sun2012intrinsic}), increasing the cavity size, and compensating losses by coherently combining outputs between stages of the cascade. Such modifications could potentially allow operation at input powers exceeding 4 W, improving the overall cascaded conversion efficiency.
 
%For example, dielectric breakdown in LiNbO$_3$~\cite{S_hickernell1972pulsed} and GaAs~\cite{S_baliga1981breakdown} occurs at over $10^7$ V/m.

% 
\begin{figure}[htbp!]
    \centering
    \includegraphics[width=0.3\textwidth]{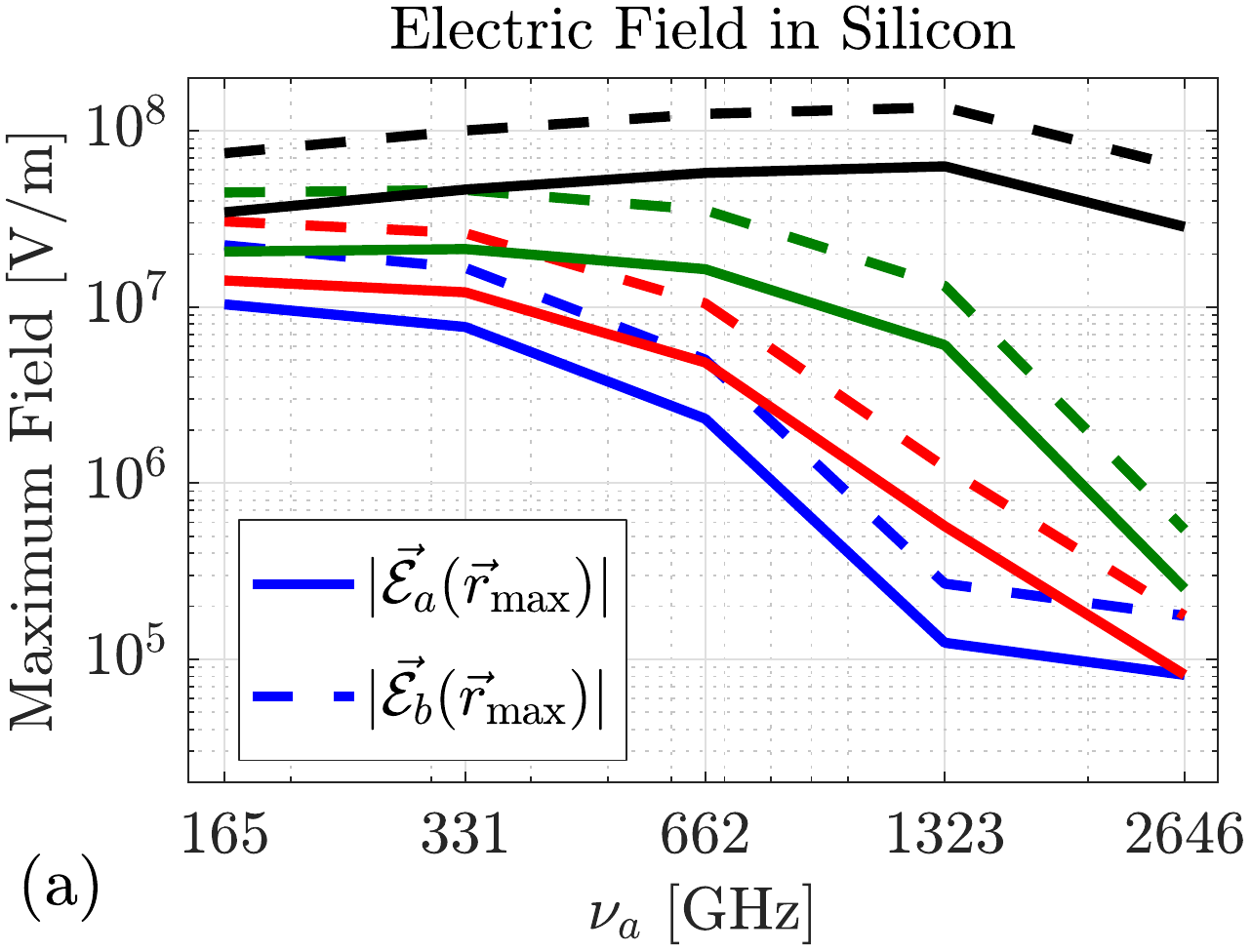}
    \hspace{4mm}
    \includegraphics[width=0.3\textwidth]{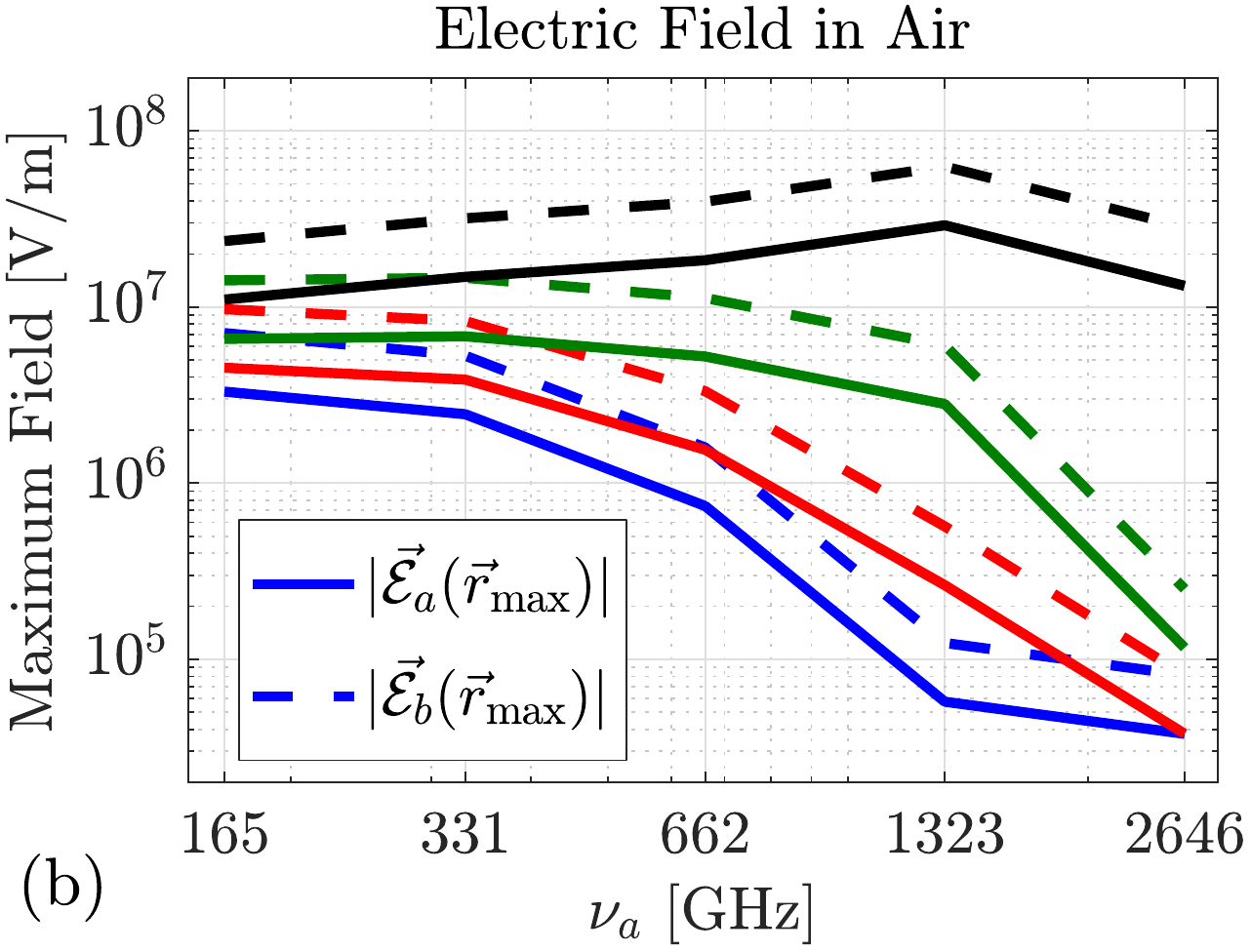}
    \hspace{4mm}
    \includegraphics[width=0.3\textwidth]{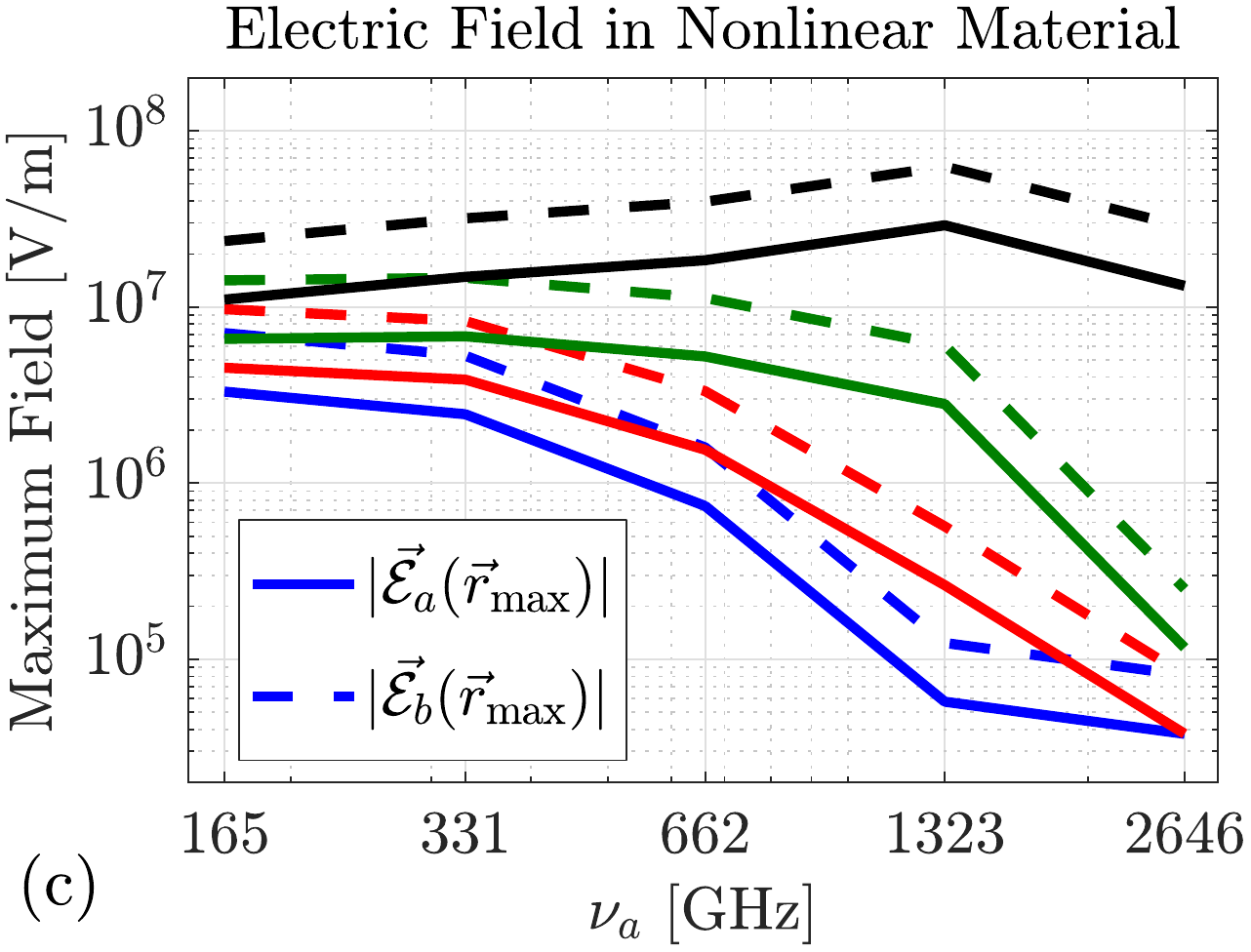}
    \caption{Maximum field strength in each of the PhC cavity materials corresponding to the power and intra-cavity energy in~\figref{phc eta abs}.} 
    \figlab{phc Emax}
\end{figure}
\begin{figure}[!htbp]
    \centering
    \includegraphics[width=0.3\textwidth]{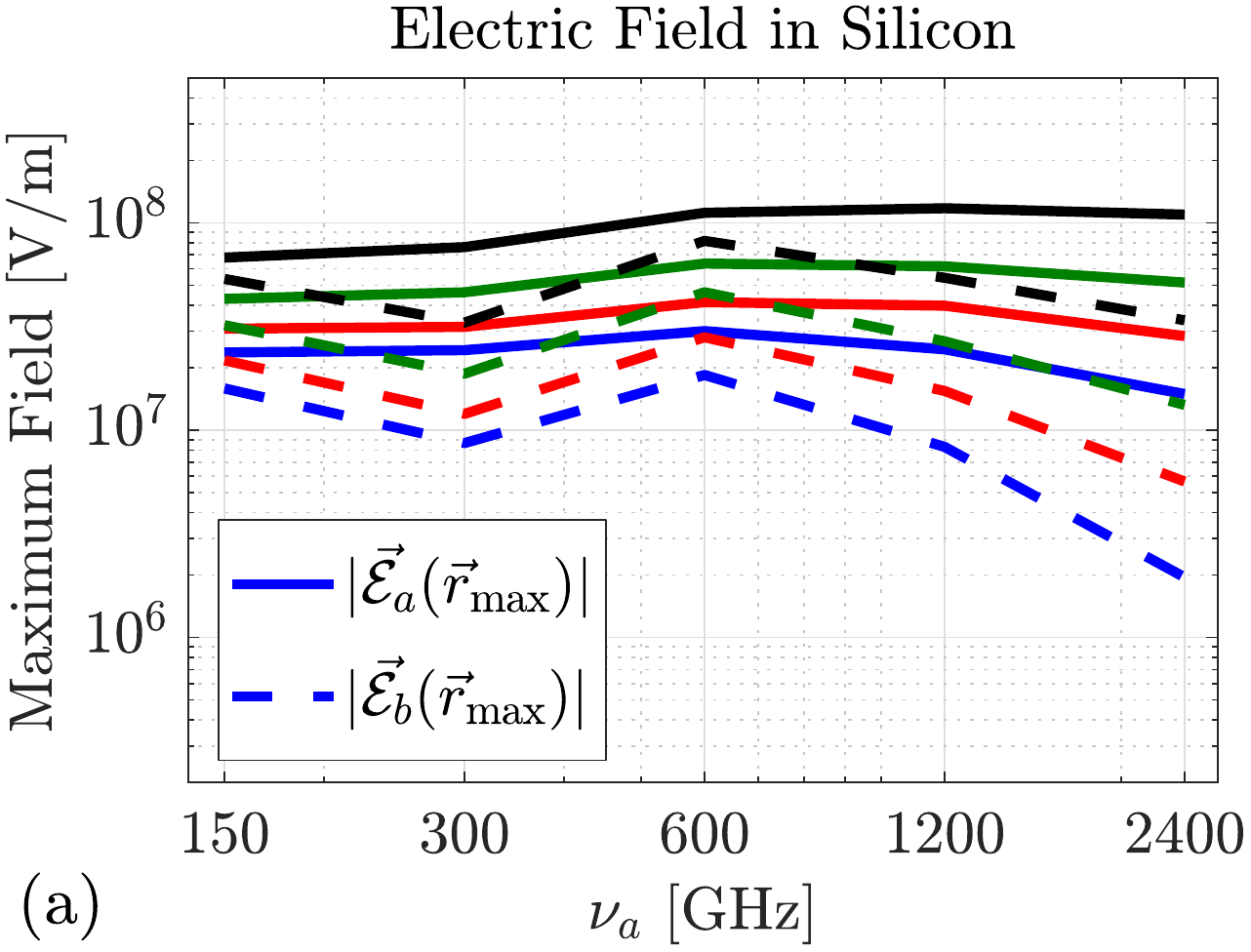}
    \hspace{4mm}
    \includegraphics[width=0.3\textwidth]{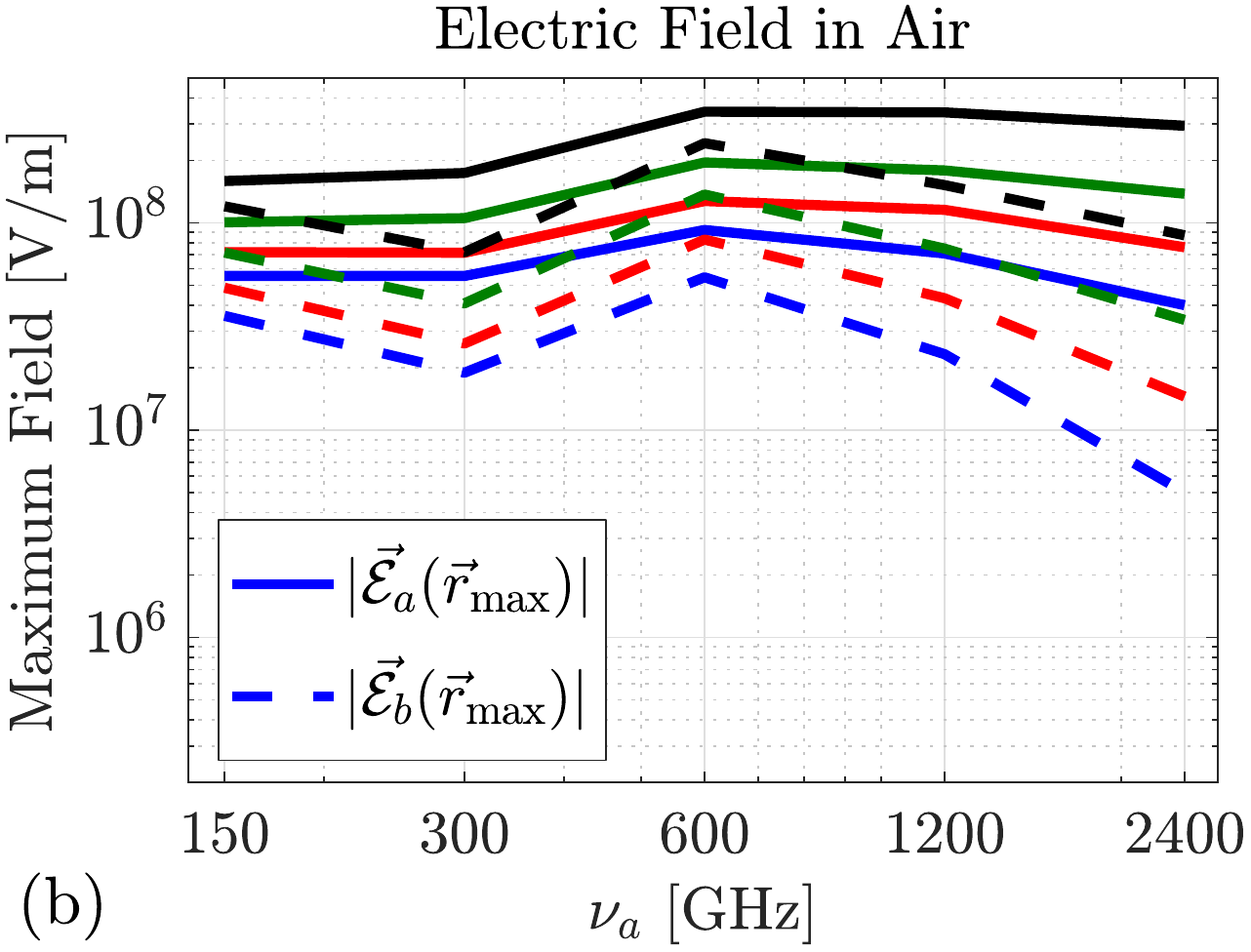}
    \hspace{4mm}
    \includegraphics[width=0.3\textwidth]{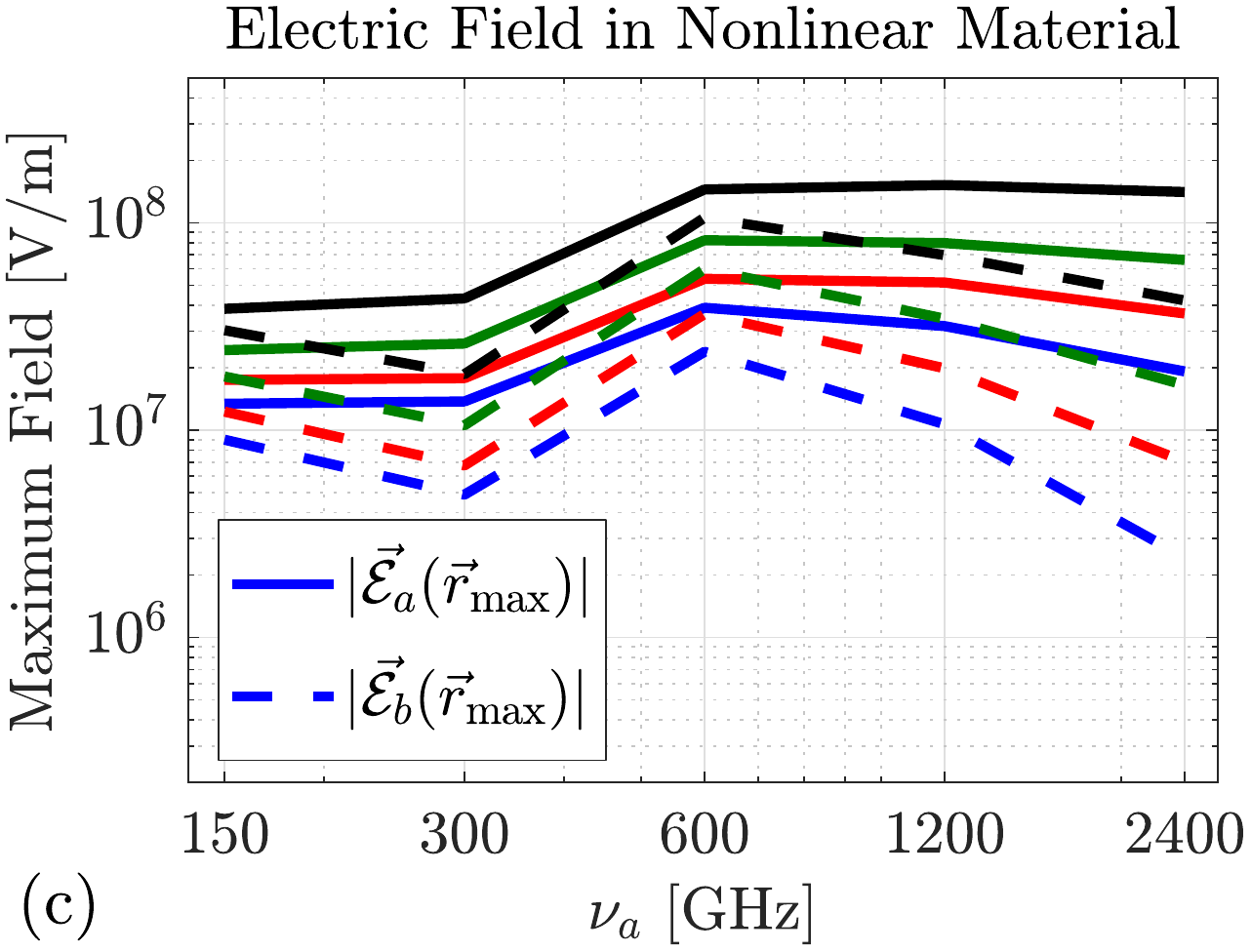}
    \caption{Maximum field strength in each of the ring cavity materials corresponding to the power and intra-cavity energy in~\figref{ring eta abs}.} 
    \figlab{ring Emax}
\end{figure}
\FloatBarrier

% Bibliography
\bibliography{supp}